\documentclass[useAMS,usenatbib]{mn2e}
\def\aap{AA}

\def\apjl{ApJL}

\def\mnras{MNRAS}
\def\apj{ApJ}
\def\apjs{ApJS}
\def\araa{ARA{\&}A}
\def\aj{AJ}
\def\pasp{PASP}

\def\nat{Nat}

\def\vd{{v_{\rm d}}}
\def\gammad{{\gamma_{\rm d}}}
\def\bin{{\beta_{\rm in}}}
\def\bout{{\beta_{\rm out}}}

\usepackage{graphicx}
\usepackage{float}
\usepackage{amssymb}
\usepackage{amsfonts}
\usepackage{amsmath} 
\usepackage{color}

\title[The GC system of M87]{Dynamical Models
 of Elliptical Galaxies -- II. M87 and its Globular Clusters} 

 \author[A. Agnello, N. W. Evans, A.J. Romanowsky, J.P. Brodie]
        {A. Agnello$^{1}$\thanks{E-mail: aagnello@ast.cam.ac.uk,nwe@ast.cam.ac.uk},
          N. W. Evans$^{1}$, A.J. Romanowsky$^{2,3}$ and
          J.P. Brodie$^{3}$ \\ $^{1}$ Institute of Astronomy,
          University of Cambridge, Madingley Road, Cambridge CB3 0HA,
          UK\\ $^{2}$ Department of Physics and Astronomy, San
          Jos{\'e} State University, One Washington Square, San
          Jos{\'e}, CA 95192, USA\\ $^{3}$ University of California
          Observatories, 1156 High Street, Santa Cruz, CA 95064, USA}

\begin{document}

\voffset-.6in

\date{Accepted . Received }

\pagerange{\pageref{firstpage}--\pageref{lastpage}} 

\maketitle

\label{firstpage}

\begin{abstract}{
 We study the Globular Cluster (GC) system of the nearby elliptical
 galaxy M87 using the newly available dataset with accurate kinematics
 provided by Strader et al. (2011).  We find evidence for three
 distinct sub-populations of GCs in terms of colours, kinematics and
 radial profiles. We show that a decomposition into three populations
 -- blue, intermediate and red GCs -- is statistically preferred to
 one with two or four populations. The existence of three components
 has been suggested before, but here we are able to identify them
 robustly and relate them to the stellar profile.  We exploit the
 sub-populations to derive dynamical constraints on the mass and Dark
 Matter (DM) content of M87 out to $\sim100$ kpc. We deploy a class of
 global mass-estimators, developed in Paper I, obtaining mass
 measurements at different locations. The DM fraction in M87 changes
 from $\approx$0.2 at the effective radius of the stellar light
 ($0.02^\circ$ or 6 kpc) to $\approx$0.95 at the distance probed by
 the most extended, blue GCs ($0.47^\circ$ or 135 kpc).

 We supplement this analysis with \textit{virial decompositions}, which exploit
 the dynamical model to produce a separation into multiple components.
 These yield the luminous mass as $5.5^{+1.5}_{-2.0} \times 10^{11} M_\odot$
 and the dark matter within 135 kpc as $8.0^{+1.0}_{-4.0} \times 10^{12} M_\odot.$
 The inner DM density behaves as $\rho \sim r^{-\gamma}$ with $\gamma\approx 1.6$.
 This is steeper than the cosmologically preferred cusp of $\rho \sim r^{-1}$
 (Dubinski \& Carlberg; Navarro, Frenk \& White), and may provide evidence
 of dark matter contraction. Finally, we combine the GC separation into
 three sub-populations with the Jeans equations, obtaining information on
 the orbital structure of the GC system. The centrally concentrated
 red GCs exhibit tangential anisotropy, consistent with the depletion
 of radial orbits by tidal shredding. The most extended blue GCs have
 an isotropic velocity dispersion tensor in the central parts, which
 becomes more tangential moving outwards, consistent with adiabatic
 contraction of the DM halo.}
\end{abstract}

 \section{Introduction}

 Elliptical galaxies and their dark matter (DM) haloes provide crucial
 tests of galaxy formation scenarios. The
 different phases of assembly \citep{joh12,hil13}, the interplay
 between luminous and dark components over different timescales
 \citep{blu86,aba10,zub12}, as well as the initial conditions
 \citep{sti87,nip12,rem13} all leave imprints on the final
 profiles. However, some properties of present-day ellipticals are not
 straightforward to obtain. For example, the orbital distribution of
 stars and the DM halo profile are not directly accessible. These must
 be derived by modelling the observed photometry and line-of-sight
 (LOS) kinematics, and are generally degenerate with one another
 (\textit{mass-anisotropy degeneracy}).

 Additional information on the mass profile can be gathered through the
 kinematics of Globular Clusters (GCs) and Planetary Nebulae
 (PNe). This has been possible thanks to dedicated observational
 campaigns and instrumentation \citep[such as the PN.Spectrograph,][]{dou02}.
 Observations of PNe probe the photometric
 and kinematic profiles of the starlight well beyond the effective
 radius, where the surface-brightness would be too low
 \citep{coc09}. Globular Cluster populations typically extend out to
 larger radii and, if their distribution and/or kinematics differ from
 those of the starlight, can be exploited as independent tracers of
 the same gravitational potential.

 Use of GC or PN kinematics as a mass constraint dates back to
 \citet{mou87} and has been applied to a diverse set of early-type
 galaxies since then \citep[e.g.][]{pot13}, to infer masses and
 properties of the DM haloes \citep{nap09,sch10,dea12} or validate
 galaxy formation scenarios \citep{coc09,sch12}. The models have relied on
 the Jeans equations or on distribution functions to reproduce the
 velocity dispersions and, in some cases, higher-order moments of the
 LOS velocity distribution. Perhaps the most complete method is given
 by orbit-based \citep[hereafter S79]{sch79} and made-to-measure modelling~\citep{sye96,mor13}.
 We will call these \textit{forward techniques}, to emphasize that they consist
 in building three-dimensional densities and velocities, which are then
 compared to the projected observables (surface density, velocity
 dispersion and higher-order moments).

 However, conclusions on the dynamical side are weakened by the
 mass-anisotropy degeneracy, such as in the controversy over DM
 fractions and orbital structure \citep{rom03,dek05}.  Also, the
 sample size and accuracy of kinematic measurements play a key
 role. For reference, reliable inference on the LOS velocity
 distribution and its higher moments requires a lower limit of
 $N\approx200$ tracers per radial bin~\citep{AmE12}, a condition which
 is not always met in practice. Especially with the limited number of
 GCs per host galaxy, only marginal conclusions on the orbital
 structure can be drawn from analyses of the higher-order moments. In
 the analogous context of resolved stellar populations in dwarf
 spheroidals (dSphs), where larger datasets are available, significant
 improvements have been brought by the identification of multiple
 populations and the use of global estimators, sometimes with sharper
 performance than more refined analyses in which all sub-populations
 were grouped together as a whole \citep{wal11,ae12,am13}. The
 virial results can then be used as a prior on the dynamical
 parameters, to inform more detailed analyses whose outcome is mainly
 the orbital structure \citep{ric13}.

 Moreover, forward techniques require a number of steps to relate the
 projected distribution on the sky to the projected kinematics. Given
 the generally low number of GCs or PNe per galaxy (a few hundred in
 the best cases, so far), the surface density distribution $\Sigma(R)$ 
 and kinematics of tracers can be very uncertain, adding to the
 computational complexity of some forward modelling techniques.
 In a companion paper \citep[see][hereafter Paper I]{a13}, a novel
 approach has been illustrated. This is based upon a reformulation
 of the Jeans equations, which yields the predicted velocity
 dispersions and their aperture-averages by means of single or double
 integrals involving $\Sigma(R)$ directly, along with a kernel that depends on
 the mass and anisotropy profiles. As a by-product, different mass
 estimators at particular locations have been derived, as well as
 characterisations of the kinematic profiles and common
 aperture-corrections \citep[cf][]{jor95,cap06}.
 Some particular cases (the virial limit and isotropic tracers in
 power-law potentials) have been previously exploited in a variety of
 problems, ranging from gravitational lensing by galaxies~\citep{aae13}
 to the DM profiles of dwarf spheroidals~\citep{am13}.

 Here, we will apply the results of Paper I to the early-type galaxy
 M87 and its GC system. This is one of the most luminous and massive
 galaxies within 20 Mpc, and hosts one of the most populous GC
 systems. It was the first target of GC spectroscopy beyond the Local
 Group \citep{mou87,mou90,huc87}. Since then, its GCs have been
 studied extensively, most notably with larger kinematic samples by
 \citet{coh97}, \citet{coh00} and \citet{han01}. Interest then turned
 to other galaxies until Strader et al. (2011, hereafter S11)
 revisited M87 with a new generation of of high resolution,
 high signal-to-noise spectroscopy, providing the
 largest, publicly available, sample in the literature. This is the
 dataset we will use here.

In Section 2, we introduce the dataset that will be used in the
subsequent dynamical analyses and summarize previous results in the
literature. In Section 3, we present a partition into sub-populations,
showing that the GC system of M87 is most likely a superposition of
three distinct (and internally homogeneous) sub-populations. In
Section 4, we use the three GC populations to obtain virial
constraints on the mass profile (luminous mass, DM fractions, DM
contraction) in a manner that is independent of any orbital
structure. In Section 5, we then apply the integral form of the Jeans
equations derived in Paper I to learn about the orbital distributions.
 We comment and draw conclusions in Section 6.
 
From now on, we adopt a systemic velocity of 1307 kms$^{-1}$
 and a distance of 16.5 Mpc for M87~\citep[see e.g.,][]{Ma09,Bi10}.
 Throughout all this paper, fits and likelihood explorations are
 performed using the Markov Chain Monte Carlo technique.

\section{M87 and its Globular Clusters}

 The S11 data were based on wide-field imaging from\footnote{Acronyms:
 CHFT, Canada-France-Hawaii Telescope; DEIMOS, DEep Imaging Multi-Object
 Spectrograph; MMT, Multiple Mirror Telescope. See S11 and references therein.}
 Subaru/Suprime-Cam and CFHT/MegaCam, with follow-up spectroscopy
 on Keck/DEIMOS and MMT/Hectospec out to 185 kpc ($\approx 0.64$
 degrees). These data
 have undergone extensive vetting for reliability and the velocities
 are accurate at the $\approx10$ kms$^{-1}$ level. Comparison to the
 older generation of data (performed by S11) was however revealing.
 Several cases of `catastrophic' errors were found, with differences
 up to $\approx 10^{3}$ km/s that seem most likely to be problems with
 the older data on a smaller telescope. Unfortunately, even a few such
 unrecognized errors can have dramatic impacts on kinematic parameters
 such as the velocity dispersion, with subsequent repercussions on
 the dynamical mass. In particular, the GC dispersion profile of M87
 was previously thought to rise rapidly with radius, but with the newer
 data it is seen to remain constant or decline (as shown below and in S11).

 The kinematics of smaller datasets of GCs at smaller radii have already
 been studied before. \citet{coh97} performed a Jeans analysis,
 assuming an isotropic velocity dispersion tensor, on 205 GCs between
 $\approx0.03$ and $0.14$ degrees (i.e. $10\lesssim R/kpc\lesssim 50$).
 They used a power-law model for the total density (i.e. luminous plus DM),
 \begin{equation}
 \rho_{\rm tot}(r)\propto r^{-\gamma}\ ,
 \end{equation}
 and inferred $\gamma\approx1.3,$ corresponding to a (total) mass profile
 $M(r)\propto r^{1.7}$. \citet{rom01} studied a set of $200$ GCs by
 means of S79 orbit modelling, finding evidence for a dark halo with
 density falling off more slowly than $r^{-2}.$ \citet{wu06} applied
 a S79 orbit-based analysis, with a power-law total density, to a sample
 of $161$ GCs between $\approx0.03$ and $0.3$ degrees. They found
 $\gamma=1.6\pm0.4,$ although the likelihood has three peaks, one of
 which is almost at $\gamma= 2.$ The energy distribution resulting
 from their modelling suggested multiple distinct components, although
 the data were not enough to confirm this hypothesis robustly.
 \citet{mur11} analysed 278 GCs up to $47$ kpc ($\approx0.15$ deg) by
 means of orbit-based methods, together with stellar kinematics for
 the starlight, as to infer the (total) mass profile over a wide radial range.
 The expected (total) mass at\footnote{Or at 35 kpc with their
 adopted distance, which is 17.9 Mpc, whereas \citet{wu06} assumed
 $\approx$16 Mpc.} 32 kpc is $\approx34\%$ higher than the one
 reported in \citet{wu06}.  Also, mass-estimates from X-rays are
 almost $50\%$ lower than the ones inferred from GCs at small radii
 \citep{mur11}.

 Here, we make use of the S11 dataset, while omitting the brightest
 objects (with magnitude $i<20$) that may be a distinct population of
 ultra-compact dwarfs (as discussed in S11 and Brodie et al. 2011). We
 also do not make use of the additional DEIMOS data from
 \citet{rom12}, since these were specifically obtained around a cold
 substructure that could bias the kinematics. In particular, the
 sample with accurate kinematic measurements (the `kinematic' sample)
 is a subset of the dataset studied in S11 (the `photometric' sample).
 \begin{figure}
         \centering
         \includegraphics[width=0.45\textwidth]{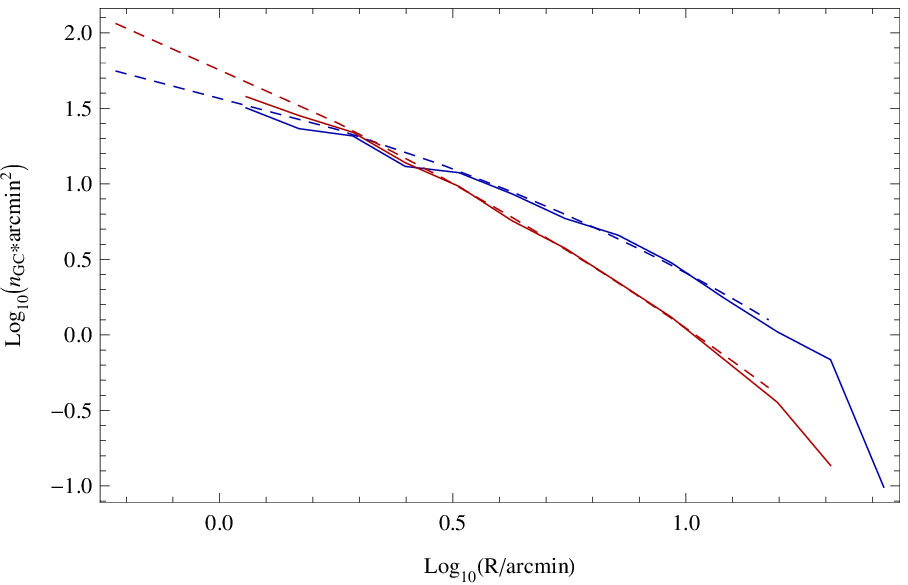}
         \includegraphics[width=0.47\textwidth]{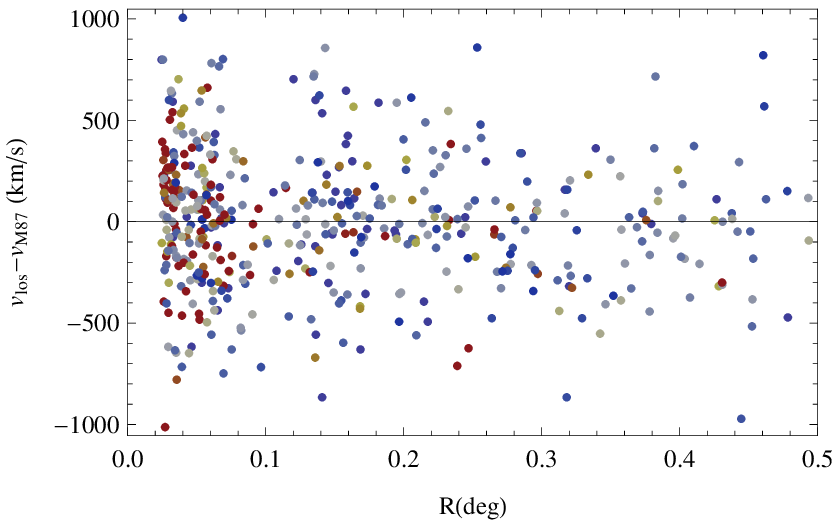}
 \caption[GCs in M87: data]{\small{ M87 Globular Cluster data.\\
     Top: Surface density profiles of GCs in M87, after
     subtraction of contaminants. The full lines are given by the
     number-counts of \citet{har09}, where the surface-density of
     contaminants is estimated directly from field objects; the dashed
     lines show the best-fitting profiles of S11, where an independent
     photometric sample was used, the fraction of contaminants was a free
     parameter in the modelling and the radial selection function was
     accounted for. Blue (dark red) denotes the GCs with $g-i$ lower
     (higher) than 0.93. Note that the redder GCs are more centrally
     concentrated.\\
     Bottom: Line of sight velocities (relative to M87) versus distance
     from M87 for GCs in the spectroscopic subsample of S11. Redder
     (bluer) points mark GCs with higher (lower) values of $g-i.$
     
     }}
 \label{fig:M87Rdist}
 \end{figure}

 \subsection{The Data}

 In order to exclude possible contaminants from our catalogue, we
 adopt the same criterion of \citet{har09} and S11, considering just
 those objects in the kinematic sample with colours as follows:
 \begin{equation}
 0.43<g\!-\!r<0.95\ ,\ \ 0.59<g\!-\!i<1.40\ .
 \end{equation}
Our kinematic sub-sample consists of $420$ bona fide GCs, with
 positions, LOS velocities and colours. A clear bimodality is present
 in the colour distribution \citep[as already observed in][S11]{har09}. 

A priori, the selection function for the kinematic sub-sample may not
 be the same as for the photometric catalogue, both in positions and in
 colours. If this happened, the surface-density profiles $\Sigma(R)$
 used in the dynamical analysis would be inadequate and any mass
 measurements would be affected by that.  Also, we must be confident
 that just the \textit{bona fide} GCs are considered for the dynamical
 analysis. \citet{har09} and S11 examined different procedures to
 isolate the GCs, subtract the contaminants and separate the sample in
 bluer ($g\!-\!i<0.93$) and redder ($g\!-\!i>0.93$) sub-populations.

 Figure \ref{fig:M87Rdist} shows the estimated radial profiles, for
 bluer and redder GCs. The profiles given in S11 are consistent with
 the number counts for pure GCs estimated by \citet{har09} with a
 different procedure for contaminant subtraction. The surface-density
 profile of `red' GCs matches smoothly to the starlight's
 surface-brightness profile in the inner regions, suggesting that the
 starlight is associated with a red GC population. This fades at
 larger radii into a bluer, metal-poorer and less concentrated
 component, which may be the result of accretion onto M87
 \citep{bro06}.

 \begin{figure}
         \centering
         \includegraphics[width=0.45\textwidth]{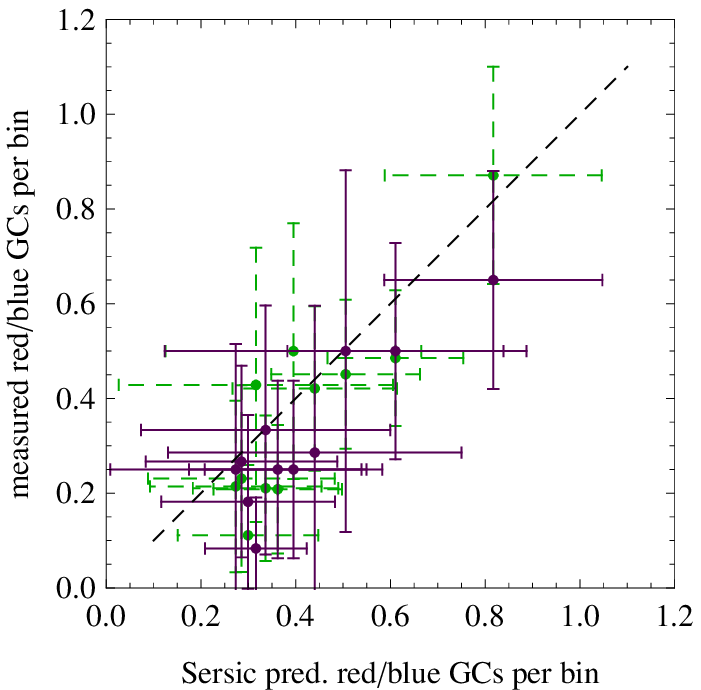}
         \includegraphics[width=0.45\textwidth]{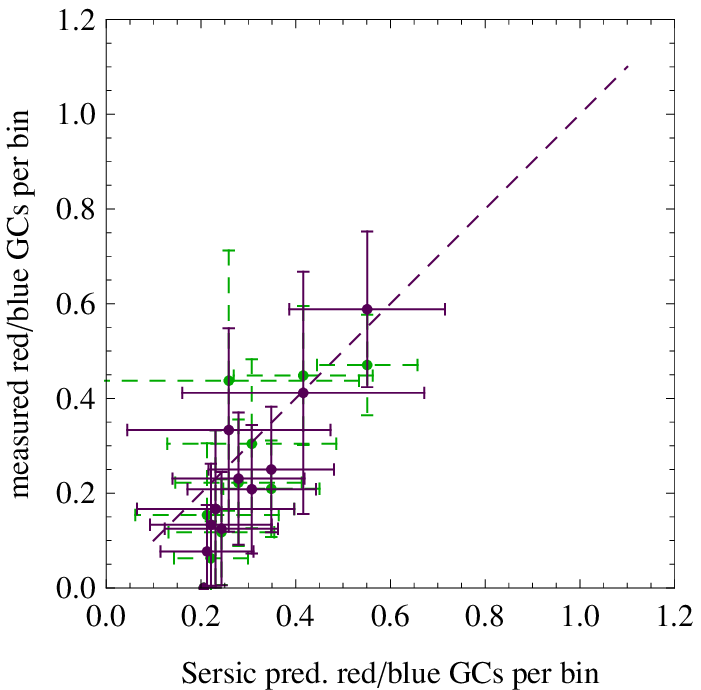}
 \caption[GCs in M87: selection effects?]{\small{ The selection
     function for the photometric and kinematic samples, see Section 2.1.
     Green dashed (purple solid) error bars denote the photometric (kinematic)
     sample. The upper panel uses bin sizes of $0.02^\circ$, the lower panel
     $0.04^\circ$. The dashed line shows the 1:1 proportionality.  The
     selection function has no appreciable variation with colour and
     has the same radial profile for both the photometric and
     spectroscopic datasets.}}
 \label{fig:M87selfnct}
 \end{figure}

 In the dynamical analysis, we will use data from the kinematic
 sub-sample. Then, we need to know if the photometric profiles are
 consistent between the photometric sample and the kinematic
 subsample. Although the radial selection function is different, a
 simple check consists in examining the ratios of redder-to-bluer GCs
 at different locations, which depends just on the colour selection
 function. Figure \ref{fig:M87selfnct} shows the ratio of
 redder-to-bluer GCs estimated by ratio of number-counts in different
 radial bins, both for the photometric and kinematic sub-samples.  As
 the plots show, the kinematic sub-sample is a faithful representation
 of the total (photometric) sample without substantial colour biases
 or contamination in the distance range probed here.

 \section{Sub-populations}

 A simple colour separation based on hard cuts \footnote{That is, $g-i$ larger or smaller
 than the conventional value 0.93.} shows that the bluer GCs have a higher
 velocity dispersion than the redder ones. Having both colours and LOS
 velocities, we can exploit them jointly to separate our sample into sub-populations.

 We will suppose that each population has a Gaussian distribution in
 colour and velocity, of the kind
 \begin{equation}
 \mathcal{G}(v,c;\sigma_{j},\langle c\rangle_{j},\Delta c_{j})=
 \frac{1}{2\pi\sigma_{j}\Delta
   c_{j}}\exp\left[-\frac{v^{2}}{2\sigma_{j}^{2}} -\frac{(c\!-\!\langle
     c\rangle_{j})^{2}}{2\Delta c_{j}^{2}}\right]\ ,
 \label{eq:gauss}
 \end{equation}
 (c.f., Walker \& Penarrubia 2011). We use the difference between
 $g$ and $i$ magnitudes as the colour $c=g-i$ in our study, while $j$
 denotes the population, $v$ the line-of-sight velocity,
 $\sigma$ the velocity dispersion and $\langle c\rangle,\Delta c$
 the mean colour and colour dispersion. The above distribution can be modified to
 distinguish between rotation and random motions. However, in what
 follows, we are primarily interested in the entirety of the second
 velocity moment, whether due to ordered or random motions, and this
 is the same as the dispersion $\sigma_{j}$ estimated via
 eq.~(\ref{eq:gauss}).

 \begin{table*}
 \begin{center}
 \begin{tabular}{| l | c | c | c | c | c | c | c | c |}
 \multicolumn{9}{c}{} \\
         \hline 
 bin & $N_{\rm bin}$ & $\langle c\rangle_{\rm r}$ & $\langle
 c\rangle_{\rm b}$ & $\Delta c_{\rm r}$ & $\Delta c_{\rm b}$ &
 $\sigma_{\rm r}$ & $\sigma_{\rm b}$ & $f_{\rm b}$ \\  
 (deg) &  &  &  &  &  & (kms$^{-1}$ ) & (kms$^{-1}$) & \\
         \hline
 $0.02<R<0.04$ & 85 & $1.05\pm0.03$ & $0.77\pm0.02$ & $0.12\pm0.02$ & $0.05\pm0.01$ & $330\pm37$ & $430\pm65$ & $0.38\pm0.09$\\
 $0.04<R<0.1$ & 123 & $1.01\pm0.04$ & $0.79\pm0.02$ & $0.11\pm0.02$ & $0.06\pm0.01$ & $265\pm29$ & $374\pm44$ & $0.51\pm0.12$\\
 $0.1<R<0.2$ & 78 & $0.92\pm0.04$ & $0.77\pm0.01$ & $0.11\pm0.02$ & $0.05\pm0.01$ & $280\pm56$ & $315\pm57$ & $0.54\pm0.11$\\
 $0.2<R<0.3$ & 68 & $0.94\pm0.04$ & $0.76\pm0.01$ & $0.12\pm0.02$ & $0.04\pm0.01$ & $295\pm48$ & $290\pm40$ & $0.59\pm0.10$\\

     \hline
 \end{tabular}
 \caption[GCs in M87: Bimodal fits]{\small{ Colour parameters,
     velocity dispersions and blue to red fractions in different
     radial bins from a bimodal fit to the colour and velocity
     distribution. Here, subscripts `r' and `b' refer to the redder
     and bluer populations respectively. The errorbars on $\langle
     g\!-\!i\rangle$ are significantly smaller than the colour
     variation with radial bin, suggesting a real change of mean
     colour (and metallicity) with radius in the redder component. The
     column $N_{\rm bin}$ indicates the number of GCs in each radial
     range.}}
 \label{tab:bimod}
 \end{center}
 \end{table*}
\begin{figure}
         \centering
         \includegraphics[width=0.45\textwidth]{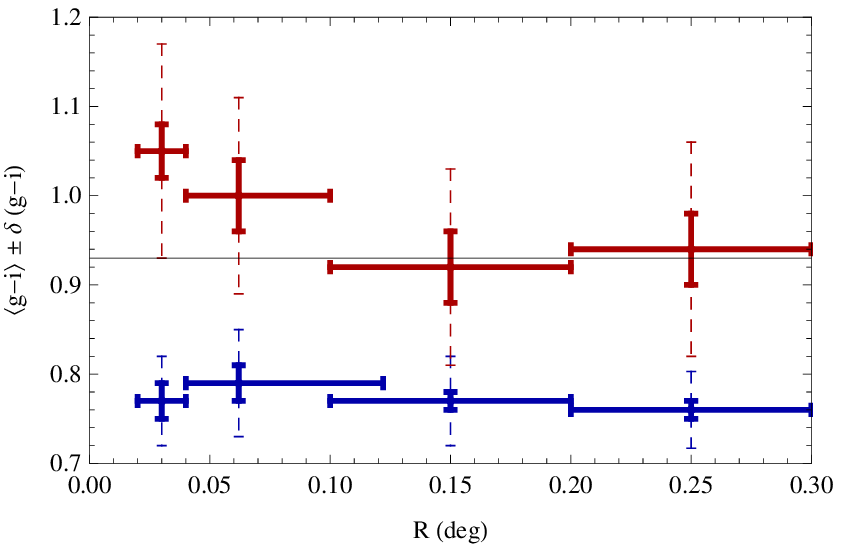}
         \includegraphics[width=0.47\textwidth]{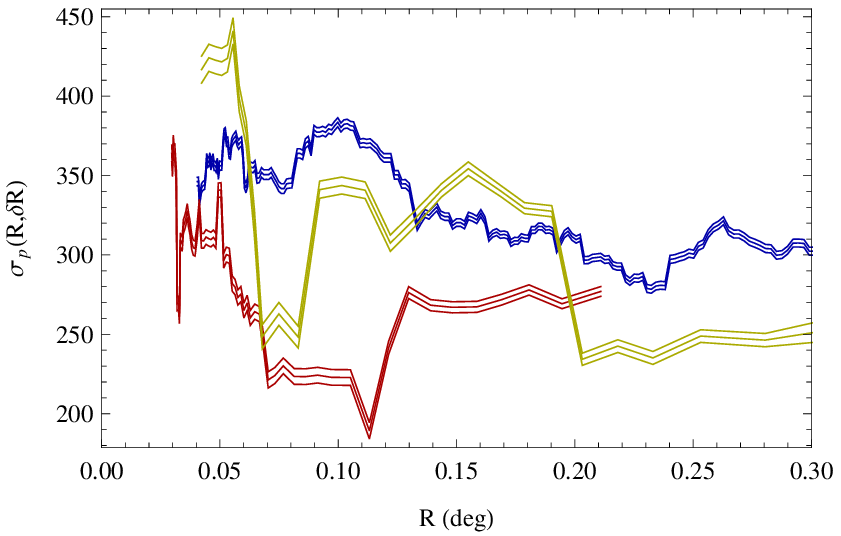}
 \caption[GCs in M87: colour and kinematic trends]{\small{Behaviour of
     GC colours and kinematics with distance from M87.\\
     Top: Colour ($g-i$) parameters resulting from a bimodal fit
     in different bins of distance (horizontal error-bars), as in Sect.3.1 and
     Table 1. Full error-bars: mean colour and its uncertainty; dashed errorbars:
     width of the colour distribution, centered on the mean colour for
     visual convenience. The mean-colour variation of the `red' GCs
     over distance is monotonic, appreciably larger than the
     uncertainties and is comparable to the colour dispersion, at
     variance with the behaviour of `blue' GCs.\\
     Bottom: Velocity dispersion profiles for GCs of different colours.
     Blue (yellow, red): GCs with $g-i<0.93$ ($0.93<g-i<1,$ $g-i>1).$
     Full lines mark the velocity dispersions computed via eq.(5), whilst
     dashed lines delimit the uncertainties computed by linear propagation
     on $v_{k}$ in eq.(5). The bin-size $\delta R$ varies with radius, such that
     a quarter of the GC subpopulation is enclosed in each bin. GCs with
     $g-i\geq 0.93$ show a sharp transition in $\sigma_{p}(R,\delta R),$
     which corresponds to the transition in mean colour of the redder GCs (top panel).
     }}
 \label{fig:M87bimod}
 \end{figure}

 \subsection{Bimodal Partition}

 We start by dividing our sample into four radial bins and performing
 a bimodal fit in each of them. We suppose that $(\sigma,\langle
 c\rangle,\Delta c)$ are reasonably uniform for each sub-population in
 each radial bin.  In the bimodal fitting, each GC with measured
 velocity $v_{k}\pm\delta v_{k}$ and colour $c_{k}\pm \delta c_{k}$
 gives a contribution
 \begin{eqnarray}
 \nonumber
 \mathcal{L}_{k}=f_{\rm
   b}\mathcal{G}\left(v_{k},c_{k};\sqrt{\sigma_{\rm b}^{2}+\delta
   v_{k}^{2}},\langle c\rangle_{\rm b},\sqrt{\Delta c_{\rm b}^{2}+\delta
   c_{k}^{2}}\right)\\ +(1\!-\!f_{\rm
   b})\mathcal{G}\left(v_{k},c_{k};\sqrt{\sigma_{\rm r}^{2}+\delta
   v_{k}^{2}},\langle c\rangle_{\rm r},\sqrt{\Delta c_{\rm r}^{2}+\delta
   c_{k}^{2}}\right)
 \end{eqnarray}
 to the likelihood of the bimodal partition, with $f_{\rm b}$ (or
 $(1-f_{\rm b})$) the fraction of `bluer' (or `redder') GCs in the
 radial bin considered. The final likelihood for the $(\sigma,\langle
 c\rangle, \Delta c)$ parameters is the product of individual
 contributions $\mathcal{L}=\prod\limits_{k}\mathcal{L}_{k}$ over all
 objects in the bin. Different possible bin sizes were investigated,
 with the aim of minimising uncertainties and bin size while keeping
 the number of GCs equal among different bins.

 Table \ref{tab:bimod} summarises the outcome of our bimodal fitting
 and the bin sizes chosen.
 The blue fraction $f_{\rm b}$ increases with radius,
 as expected based on \citet{har09} and S11. The `bluer'
 sub-population has overall constant colour parameters and a smoothly
 declining velocity dispersion profile.  The `redder' GCs have a mean
 colour that decreases towards $\langle g-i\rangle\approx 0.93.$ Also,
 their LOS kinematics shows an abrupt drop-off from $\approx 330$
 kms$^{-1}$ in the innermost bin to $\approx 270$ kms$^{-1}$ at larger radii.
 This is accompanied by a significant change in the mean colour of
 the `redder' sub-population. The inner value of the velocity dispersion is consistent
 with the kinematics of the stellar component (S11). Suppose the
 velocity dispersion of `redder' GCs is estimated at different radii
 via a uniform smoothing kernel
 \begin{equation}
 \sigma^{2}_{\rm p}(R;\delta r)=\frac{1}{N}\sum\limits_{k=1}^N
 v_{k}^{2}\,\mathbf{1}_{|R_{k}-R|<\delta r}\ ,
 \end{equation}
 where $\mathbf{1}$ is the Heaviside function. Then, the dispersion
 still shows a sharp decrease for the `redder' GCs, which does not have
 a counterpart in the `bluer' ones (fig.\ref{fig:M87bimod}).
 On the other hand, the LOS kinematic profile
 for the starlight is approximately
 flat even beyond the effective radius, so we would not expect
 a sharp drop in $\sigma_{p}(R)$ for the red GCs either.

 The following hypothesis emerges: the red GCs consist of a more
 compact (and redder) sub-population and a more extended,
 intermediate-colour sub-population. This hypothesis was suggested and
 supported also in Sections 5 and 6 of S11. Our sample is a
 subset\footnote{ Besides those used here, S11 included other GCs,
   with measurements already known from the literature but with
   unverified velocities.}  of the one considered by S11, which had
 been thoroughly examined to exclude contaminants or evident
 substructure and select just the GCs that are bound to M87. Here, we
 will make one further step and exploit this hypothesis to identify
 three dynamical tracer populations, which can be used to infer masses
 at different locations.

 \begin{table}
 \begin{center}
 \begin{tabular}{| l | c | c | c |}
 \multicolumn{4}{c}{} \\
         \hline 
 parameter & blue & intermediate & red \\  
         \hline

 $\langle c\rangle$ & $0.773\pm0.005$ & $0.93\pm0.03$ & $1.10\pm0.02$ \\
 $\Delta c$ & $0.047\pm0.004$ & $0.09\pm0.01$ & $0.074\pm0.009$ \\
 $\sigma$ (km/s) & $350\pm19$ & $275\pm25$ & $333\pm37$ \\
 $n$ & $2.25^{+1.45}_{-0.45}$ & $4.0^{+1.1}_{-1.8}$ &
         $4.68^{+0.90}_{-1.10}$ \\
 \null & \null & \null & \null \\
 \multicolumn{4}{c}{$\Sigma_{\rm i/
     b}=20^{+10}_{-7},$ $\Sigma_{\rm r/b}=2480^{+3120}_{-1060}$} \\
 $R_{\rm e}(\rm{arcsec})$ & $2300\pm 700$ & $1750 \pm 650$ & $72^{+128}_{-22}$ \\
         \hline\hline
 $\langle g-i\rangle$ & $0.773\pm0.004$ & $0.92\pm0.02$ & $1.09\pm0.03$  \\
 $\Delta c$ & $0.047\pm0.003$ & $0.09\pm0.02$ & $0.08\pm0.02$ \\
 $\sigma$ (km/s) & $350\pm15$ & $267\pm12$ & $326\pm34$ \\
 \null & \null & \null & \null \\
 \multicolumn{4}{c}{$f_{\rm b}=0.54\pm0.05,$ $f_{\rm i}=0.34^{+0.07}_{-0.13}$} \\
         \hline
   \end{tabular}
 \caption[GCs in M87: three populations]{\small{ The dissection into
     three populations with (upper) and without (lower) positional
     information. In the upper table, we give colour parameters, mean
     velocity dispersions and S{\'e}rsic indices for the three
     sub-populations. The remaining photometric parameters are not
     well constrained, so underneath we list the effective radii and
     normalisations once $\langle g-i\rangle_{j},$ $\delta c_{j},$
     $\sigma_{j}$ and $n_{j}$ are fixed to their maximum likelihood
     values.  The posterior on effective radii has a large width (cf
     S11), especially for the blue and intermediate populations.  In
     the lower table, we list the parameters of a separation into
     three components in which no information from the positions is
     used. Reassuringly, the best-fit colour parameters and velocity
     dispersions do not change significantly. Also, the parameters
     $R_{\rm e,r},R_{\rm e,i},$ $\Sigma_{\rm{i/b}}/\Sigma_{\rm{r/b}}$
     are very close to the ones derived by \citet{jan10} on the
     starlight.}}
 \label{tab:3pop}
 \end{center}
 \end{table}

 \subsection{Three Populations}
 \label{sect:3pops}

 We proceed to a decomposition into three sub-populations, each of
 which has uniform colour parameters and velocity dispersion.  We model
 each sub-population with a S{\'e}rsic profile for the surface number-density
 \begin{equation}
 \Sigma_{j}(R)=\Sigma_{0,j}\exp\left[-\kappa_{n_{j}} \left(R/R_{{\rm e},j}\right)^{1/n_{j}}\right]\ ,
 \end{equation}
 and denote the fraction of $j-$th population at radius $R$ as 
 \begin{equation}
 f_{j}(R)=\frac{\Sigma_{j}(R)}{\Sigma_{\rm b}(R)+\Sigma_{\rm
     r}(R)+\Sigma_{\rm i}(R)},
 \end{equation}
where we have added a new subscript `i' to denote the
 intermediate population.  The S{\'e}rsic coefficient $\kappa_{n}$ is
 chosen such that $R_{{\rm e},j}$ is the effective radius of the
 $j-$th population, enclosing half the total (projected) number of GCs.
 In particular, we rely on the approximate expression listed
 in \citet{cio99} in terms of the S{\'e}rsic index $n.$

Then, a GC at radius $R_{k}$ will give a contribution to the likelihood
 \begin{eqnarray}
 \nonumber \mathcal{L}_{k}=f_{\rm
   b}(R_{k})\mathcal{G}\left(v_{k},c_{k};\sqrt{\sigma_{\rm
     b}^{2}+\delta v_{k}^{2}},\langle c\rangle_{\rm b},\sqrt{\Delta
   c_{\rm b}^{2}+\delta c_{k}^{2}}\right)\\
 \nonumber +f_{\rm
   r}(R_{k})\mathcal{G}\left(v_{k},c_{k};\sqrt{\sigma_{\rm
     r}^{2}+\delta v_{k}^{2}},\langle c\rangle_{\rm r},\sqrt{\Delta
   c_{\rm r}^{2}+\delta c_{k}^{2}}\right)\\
 +f_{\rm i}(R_{k})\mathcal{G}\left(v_{k},c_{k};\sqrt{\sigma_{\rm i}^{2}+
\delta v_{k}^{2}},\langle c\rangle_{\rm i},\sqrt{\Delta c_{\rm i}^{2}+\delta c_{k}^{2}}\right)\ .
 \label{eq:like3pop}
 \end{eqnarray}
 Just as for the bimodal fits, the likelihood of the parameters is the
 product of $\mathcal{L}_{k}$ over all GCs in our sample. The
 likelihood is now a function of: effective radii $R_{\rm e}$ and
 S{\'e}rsic indices $n$ of the populations; mean colours and colour
 dispersions; global velocity dispersions; and the ratios $\Sigma_{\rm
   r/b}=\Sigma_{\rm r}(0)/\Sigma_{\rm b}(0),$ $\Sigma_{\rm
   i/b}=\Sigma_{\rm i}(0)/\Sigma_{\rm b}(0)$ between the surface number
 densities at the centre. By studying the likelihood over the parameter
 space, we obtain both the best-fitting parameters and their distribution,
 including uncertainties.

 Table \ref{tab:3pop} shows the colour parameters, mean velocity
 dispersions and photometric parameters.  Uncertainties on the
 photometric parameters are large, especially for the effective
 radii. This was also observed in S11, where photometric profiles were
 sought for the bluer ($g-i<0.93$) and redder ($g-i>0.93$) GCs in the
 larger, photometric sample. Despite the uncertainties, some
 well-defined relations are satisfied among the parameters. In
 particular, the effective radius of the intermediate population is
 almost always smaller than that of the blue GCs (with $P(R_{\rm
   e,i}>R_{\rm e,b})<10\%$).  Also, the velocity dispersions are
 strictly ordered as $\sigma_{\rm b}>\sigma_{\rm r}>\sigma_{\rm i}$
 ($P(\sigma_{\rm i}>\sigma_{\rm r})< 10\%$ and $P(\sigma_{\rm
   r}>\sigma_{\rm b})< 4\%$).

 We can also examine a fit with three components in which no
 information on the radial positions is used. This is a way of
 checking that the findings on colours and kinematics are robust. In
 this case, all GCs are grouped together in the same distance bin and
 the fractions $f_{b},$ $f_{\rm i},$ $f_{r}=1-f_{b}-f_{\rm i}$ are
 free parameters themselves. The result is shown in the bottom section
 of Table~\ref{tab:3pop} and confirms the findings based on radial
 profiles.

 We also investigate a different choice of velocity dispersion law,
 which can be rising or falling at large radii (cf Table 1), namely\footnote{
 With a spatially varying velocity dispersion, the likelihood is simply
 modified by replacing $\sigma_{j}$ with $\sigma_{j}(R_{k})$ for the $k-$th
 GC.
 }
 \begin{equation}
 \sigma_{\rm p}(R)=\sigma_{0}+\frac{\sigma_{1}R_{\rm
     a}^{\zeta}}{(R_{\rm a}^{2}+R^{2})^{\zeta/2}}\ .
 \label{eq:sprofblue}
 \end{equation}
 Use of equation (\ref{eq:sprofblue}) for the kinematics of any of the
 subpopulations does not improve the likelihood of the fit appreciably,
 as the increase in $\log\mathcal{L}$ is not sufficient to balance the
 increase in degrees of freedom. Moreover, the global velocity
 dispersion of each population $\sqrt{\langle\sigma^{2}_{{\rm
       p},j}\rangle}$, computed from
 \begin{equation}
 \langle\sigma^{2}_{\rm
   p}\rangle=\frac{\int_{0}^{\infty}R\Sigma(R)\langle \sigma^{2}_{{\rm
       p},j}\rangle(R)\mathrm{d}R}{\int_{0}^{\infty}R\Sigma(R)\mathrm{d}R}\ ,
 \end{equation}
 does not differ appreciably from the result of a fit with uniform
 $\sigma.$ In other words, the fit deploying a parameterization of
 $\sigma$ yields some marginal information on the velocity dispersion
 profiles, without changing the global velocity dispersion of each
 population. The effective radii $R_{\rm{e,r}},$ $R_{\rm{e,i}}$ and
 the surface-density ratio
 $\Sigma_{\rm{i/b}}/\Sigma_{\rm{r/b}}=\Sigma_{\rm{i}}(0)/\Sigma_{\rm{r}}(0)$
 are remarkably close to the findings of a double de Vaucouleurs fit
 to the starlight, as performed by \citet{jan10}. This lends further
 support to the robustness of our results and suggests a link between
 GC populations and the stellar component.

 \begin{table}
 \begin{center}
 \begin{tabular}{| l | c | c |}
 \multicolumn{3}{c}{} \\
         \hline 
 number of & $\log\mathcal{L}$ & $\log_{10} Z$\\
 populations
  & \null & \null \\
         \hline
 2 & $-$2694.0 & $-$1177.4\\
 3 & $-$2691.8 & $-$1175.4\\
 4 & $-$2691.0 & $-$1175.2\\
         \hline
 \end{tabular}
 \caption[GCs in M87: Statistical evidence for three populations]{\small{
     Partitions using information on colours and
     velocities, but not on distances from the center of M87. The
     maximum likelihood of each decomposition is listed
     ($\log\mathcal{L}$) as well as the logarithmic evidence
     ($\log_{10} Z$).}}
 \label{tab:bayes}
 \end{center}
 \end{table}

An important fact is the independence of these findings from any
dynamical model for M87 and its gravitational potential.  Then, for a
dynamical model to be reliable, it must be able to reproduce the main
features of this decomposition.  A criterion to assess the
acceptability of a dynamical model will be whether the ordering
$\sigma_{\rm b}>\sigma_{\rm i}>\sigma_{\rm r}$ in velocity dispersions
and $R_{\rm e,b}>R_{\rm e,i}$ in effective radii are satisfied.

 \subsection{Statistical Evidence for Three Populations}

 The separation into three components gives a smoother description of
 the GC system of M87. To ensure that this choice is indeed preferable,
 we must quantify how well it describes the coulours and kinematics
 with respect to a partition into two sub-populations. We consider two
 different \textit{Bayes factors} \citep{Je61, ken79, bur02} as quantitative
 criteria, namely the ratio of maximum likelihoods and of evidences.

 Given two models A and B with maximum likelihoods $\mathcal{L}_{\rm
   A}>\mathcal{L}_{\rm B},$ let us define
 \begin{equation}
 \Delta\chi^{2}=2\log\mathcal{L}_{\rm A}-2\log\mathcal{L}_{\rm B}.
 \end{equation}
 In the case of least-square fitting with Gaussian statistics, this
 would be exactly the decrease in $\chi^{2}.$ Now let the number of
 free parameters in $A$ exceed the one in $B$ by an amount $\Delta p.$
 Then, the larger the difference
 \begin{equation}
 B_1= \Delta\chi^{2}-\Delta p,
 \label{eq:bayes1}
 \end{equation}
 the more model $A$ is preferable to model $B.$ Equation
 (\ref{eq:bayes1}) then defines the first Bayes factor, $B_1$.

 On the other hand, by introducing a new set of parameters, there is a
 risk of over-fitting. To avoid a fine-tuning of the new parameters to
 reproduce the observations, it is useful to examine the evidence
 ratio
 \begin{equation}
 \frac{Z_{\rm A}}{Z_{\rm B}}\equiv\frac{P({\rm A|data})}{P({\rm B|data})}=
 \frac{\int\mathcal{L}(\mathbf{p}_{\rm A})\theta(\mathbf{p}_{\rm
     A}|M_{\rm A})\mathrm{d}\mathbf{p}_{\rm A}}
 {\int\mathcal{L}(\mathbf{p}_{\rm B})\theta(\mathbf{p}_{\rm B}|M_{\rm
     B})\mathrm{d}\mathbf{p}_{\rm B}}\ ,
 \end{equation}
 where $\theta$ indicates the priors and $\mathcal{L}$ the likelihoods
 in the two models $M_{A},M_{B}.$  In particular, we will quote the results in terms
 of
 \begin{equation}
 B_2 = \Delta\log_{10}Z=\log_{10}(P({\rm A|data})/P({\rm B|data}))\
 \label{eq:bayes2}
 \end{equation}
 as the second Bayes factor.

 Table~\ref{tab:bayes} shows the results for a decomposition into two,
 three or four sub-populations, where no information from the positions
 of the GCs is used. Each additional population increases the number of
 free parameters by 4 ($f_i, \sigma_i, \Delta c_i$ and $\langle c
 \rangle_i$). On moving from two populations to three,
 $\Delta\chi^{2}=2\Delta\mathcal{L}=4.4$ so that the first Bayes factor
 $B_1$ indeed indicates that a separation into three populations is
 preferred over two. On moving from three populations to four,
 $\Delta\chi^{2}=2\Delta\mathcal{L}=1.6$, which does not outweigh the
 increase in the numbers of free parameters.  This can be confirmed by
 examining the evidence, using the second Bayes factor $B_2$. On
 comparing the hypothesis of three populations against two, $B_2 = 2.0$
 so that the evidence in favour of three populations is `very strong'
 on the \citet{Je61} scale. By contrast, the evidence in favour of
 four populations over three is `barely worth mentioning' ($B_2 =
 0.1$).

 Another way of testing if three populations are actually a better fit
 is to consider the gain in the first Bayes factor arising when
 information on the radial positions is included. Now the two
 population likelihood is $\mathcal{L}=-2673$ whilst the three
 population likelihood is $\mathcal{L}=-2662$. With an increase $\Delta
 p=6$ (a S{\'e}rsic index and scalelength in addition to the previous listed
 4 parameters) and a gain $\Delta\chi^{2}=2\Delta\mathcal{L}=22,$ we can
 be confident that the GC system of M87 is indeed the superposition of
 three sub-populations.

 \begin{figure}
         \centering
         \includegraphics[width=0.45\textwidth]{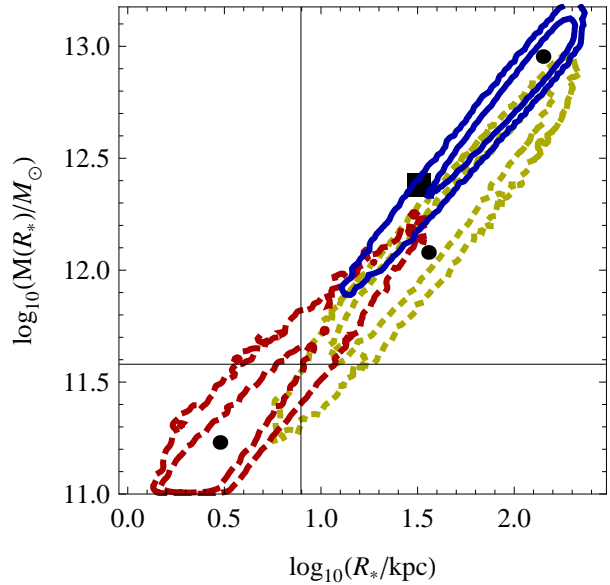}
 \caption[GCs in M87: virial pinch radii and enclosed masses]{\small{
     Distribution of the virial pinch radius $R_{\star}$ and of the (total) mass
     enclosed within it, using eq.s~(16) and (\ref{eq:inourcase}). The
     distribution of $R_{\star}$ and $M(R_{\star})$ is
     obtained from the posterior on the parameters in the
     three-population dissection. Blue and solid (yellow and dotted,
     red and dashed respectively) contours are the $68\%$ and $95\%$
     confidence regions as given by the blue (intermediate, red)
     population. The axes intersect at the value estimated from the
   starlight. A square marks the value found by \citet{wu06}.}}
 \label{fig:mest}
 \end{figure}
 \begin{table}
 \begin{center}
 \renewcommand{\tabcolsep}{0.2cm}
 \renewcommand{\arraystretch}{0.5}
 \begin{tabular}{| l | c | c | c | c | c |}
 \multicolumn{6}{c}{} \\
         \hline 
 pop. & $R_{\star}$ & $M(R_{\star})$ & $v_{\rm c}(R_{\star})$ & $M/L$ & $f_{\rm d}$\\  
  & (deg) & $(M_{\odot})$ & (km/s) & $(M/L)_{\odot,V}$ & \\ 
         \hline
 red & 0.01 & $1.7\times 10^{11}$ & 500 & 17.0 & 0.1\\
 int. & 0.10 & $1.2\times 10^{12}$ & 420 & 31.6 & 0.55 \\
 blue & 0.47 & $9.0\times 10^{12}$ & 530 & 129.0 & 0.94 \\
         \hline
 starlight & 0.02 & $3.4\times 10^{11}$ & 500 & 22.7 & 0.2\\
     \hline
   \end{tabular}
 \caption[GCs in M87: virial pinch radii and enclosed masses]{\small{
 Masses enclosed at the pinch radii of the blue,
     intermediate and red populations, together with implied circular
     velocities. The quoted values of $R_{\star}$ correspond to the
     peak of the posterior distribution in logarithmic bins. The
     quoted value of mass corresponds to the peak of the distribution
     in $M(R_{\star})$, inherited from the distribution of
     $\sigma_{j},$ $\log R_{{\rm e},j}$ and $n_{j}$ via
     eq.~(\ref{eq:inourcase}). The V-band mass-to-light ratio
     is computed adopting the double de Vaucouleurs fit to the
     starlight described by~\citet{jan10}. The final column gives the dark matter
     fraction, using the luminosity profile discussed in~\citet{Mc99},
     although comparable results are obtained with different parameterisations
     \citep[such as in][]{rom01}.
     }}
 \label{tab:masses}
 \end{center}
 \end{table}

 \section{Dynamics}

 Multiple stellar populations can provide a powerful constraint on the
 underlying gravity field. This is because each population must be in
 dynamical equilibrium in the same gravitational potential. A number of
 authors have exploited this fact to constrain the mass distribution,
 particularly in dwarf spheroidal galaxies (e.g., Walker \& Penarrubia
 2011, Amorisco \& Evans 2012). For the sake of comparison, the analyses
 of multiple populations in dwarf spheroidals could rely on datasets with
 more than $10^{3}$ stars each, whereas here we have just $420$
 confirmed GCs at our disposal.

 \subsection{Masses within Pinch Radii}

 Before embarking on elaborate models, we begin with something simple.
 For any stellar distribution with velocity dispersion $\sigma$ and
 effective radius $R_{\rm e}$, Paper I shows how to identify a pinch
 radius $R_\star$, within which the uncertainty in the (total) enclosed mass
 is minimised. We summarise the method here for the reader's convenience.

 When the kinematics of a tracer population are probed over large
 radii\footnote{The necessary radial coverage is better discussed
 in Paper I. Here it suffices to know that the necessary requirements
 for these virial analyses are met. This also results a posteriori
 from the Jeans analysis of Section 4.4, which essentially adds further
 insight on the GC orbits but not on the mass profile.},
 the mass within the pinch radius is:
 \begin{equation}
 M(R_{\star})= K
 \frac{R_{\star}\langle\sigma^{2}\rangle}{G}\ ,\qquad\qquad K\approx
 2.3.
 \label{eq:paperi}
 \end{equation}
 The proportionality coefficient $K \approx 2.3$ is an average over
 different mass models.  For a S{\'e}rsic profile of effective radius
 $R_{\rm e}$ and index $n$, the pinch radius is
 \begin{equation}
 R_{\star}=R_{e}\kappa_n^{-n}\sqrt{2\Gamma(3n)/\Gamma(n)}\ .
 \label{eq:paperia}
 \end{equation}
  The relation~({\ref{eq:paperi}) is more conveniently restated in our case as
 \begin{eqnarray}
\nonumber M(R_{\star,j}) &=& 1.54\left(\frac{\sigma_{j}}{100\ \mathrm{km/s}}\right)^{2}\left(\frac{R_{\star,j}}{\mathrm{deg}}\right)\times10^{12}M_{\odot}\\
 &=& 0.53\left(\frac{\sigma_{j}}{100\ \mathrm{km/s}}\right)^{2}\left(\frac{R_{\star,j}}{10\mathrm{kpc}}\right)\times10^{11}M_{\odot}
 \label{eq:inourcase}
 \end{eqnarray}
 for the $j-$th population (i.e. blue, red or intermediate-colour).
 For multiple populations, such formulae can be applied to
 each population, giving insights into the variation of the mass with
 radius. In our case, the values of $R_{j},\sigma_{j},n_{j}$ are given by
 the MCMC likelihood exploration in the partition into sub-populations.

 The results for the three sub-populations are given in Table
 \ref{tab:masses} and visualised in Fig.~\ref{fig:mest} \citep[see][for
 a similar approach]{wal11,am13}. We also quote the pinch
 radius and enclosed mass for the starlight, simply described by a de
 Vaucouleurs profile with effective radius $R_{e}=0.02$ degrees
 \citep{har09} and velocity dispersion $\sigma_{p}\approx 330$
 km/s (see S11 and references therein). These simple estimates
 suffer from the systematics that we have illustrated in Paper I. In
 particular, they do not account for the distribution of $K$ given by
 the (unknown) mass profile. Notice that the likelihood contours in
 Fig.~\ref{fig:mest} are distended along the line $M(R) \propto R$,
 which corresponds to isothermal. This is because the relative uncertainties
 in the effective radii are larger than those in $\sigma_{j},n_{j}$
 and the use of formulae akin to eq.~({\ref{eq:paperi}, \ref{eq:paperia})
 causes this to propagate linearly into the uncertainty on the enclosed mass.
 Barring systematic uncertainties (see Paper I), the total mass enclosed
 within the most likely pinch radius has a relative uncertainty of
 $\approx 0.1$ dex for each population.

 \begin{figure}
         \centering
         \includegraphics[width=0.45\textwidth]{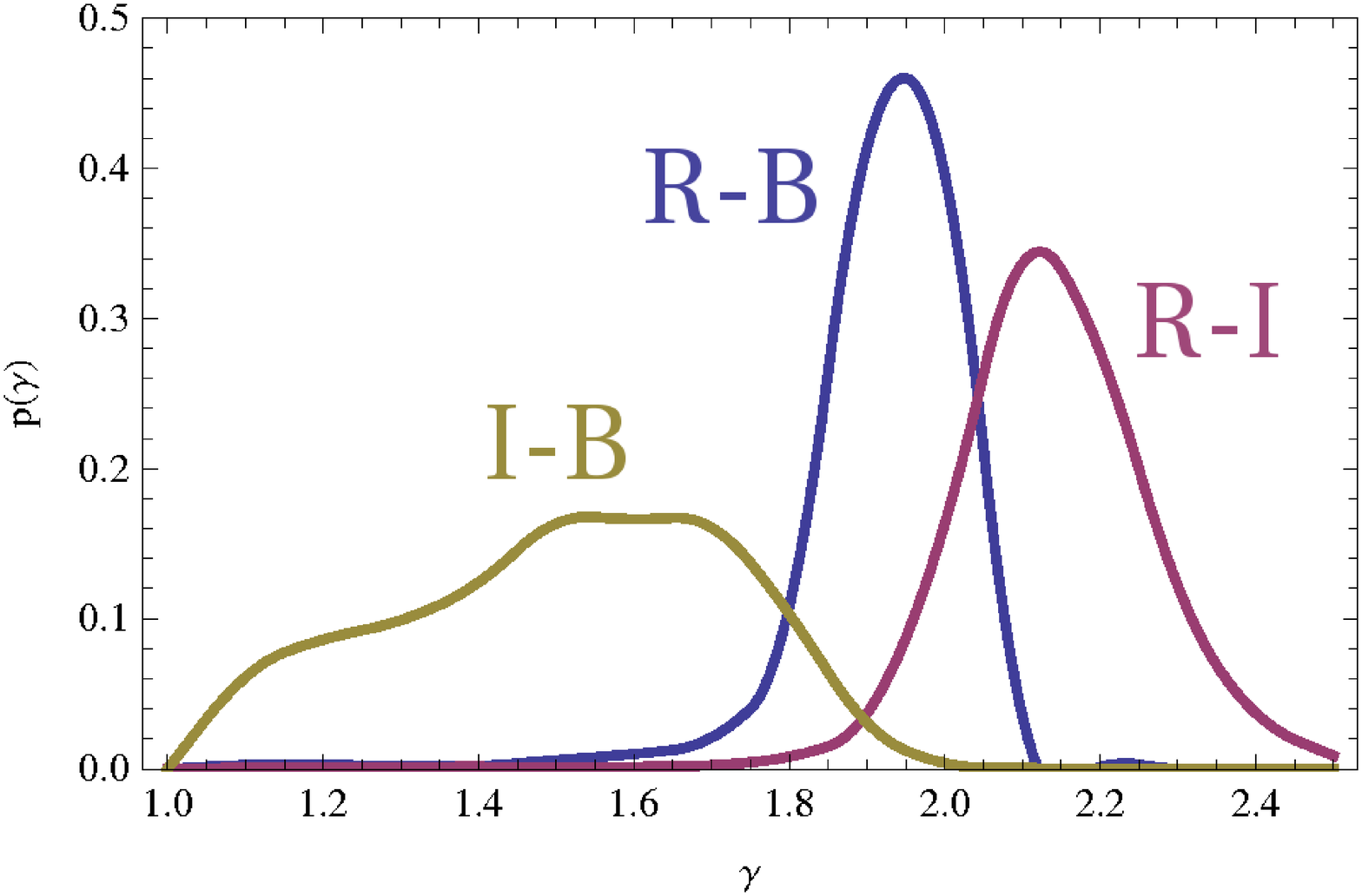}
         \includegraphics[width=0.45\textwidth]{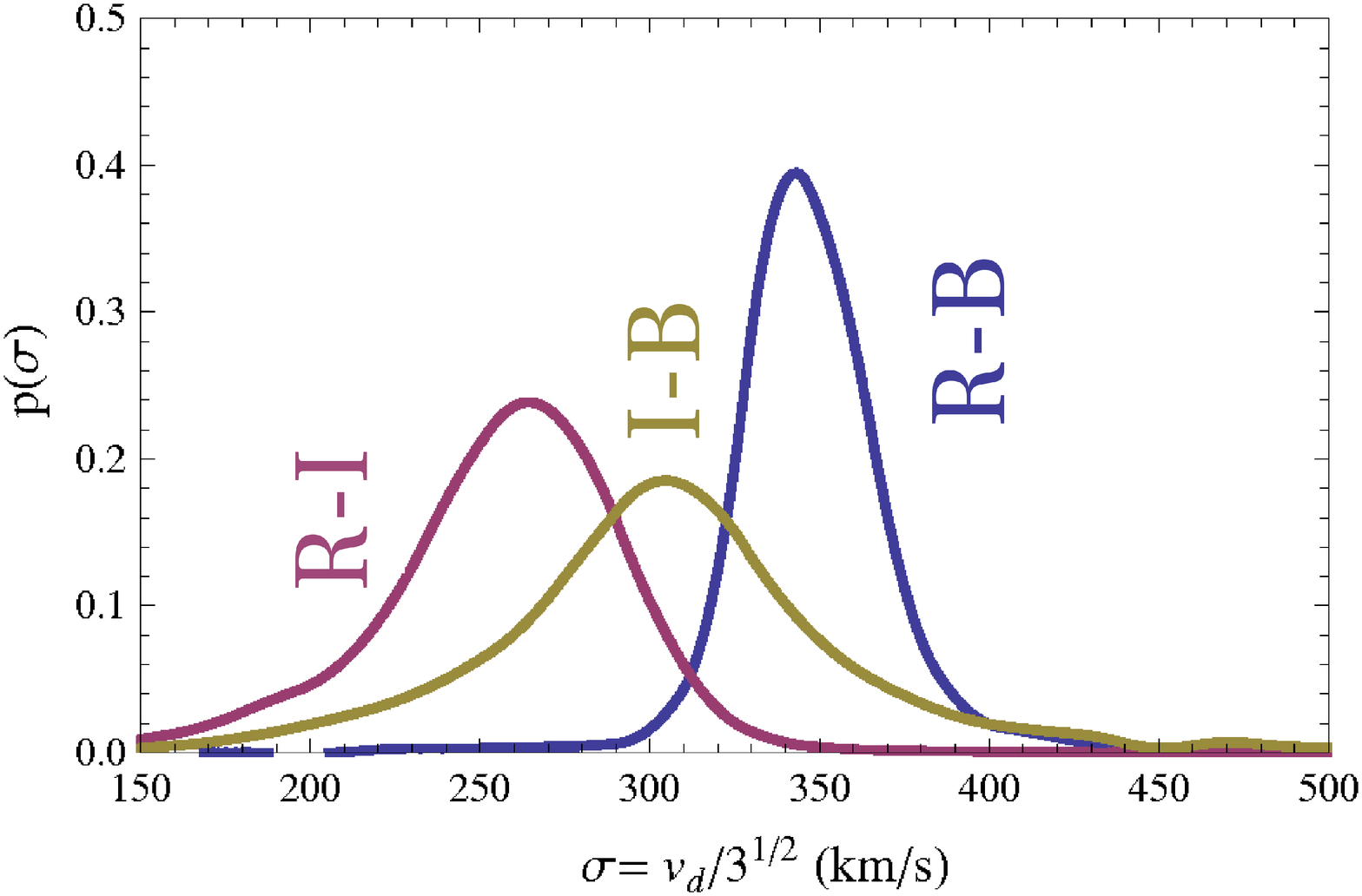}
 \caption[GCs in M87: power-law total density]{\small{ Marginalised
     likelihood for the density exponent (top) and velocity
     normalisation (bottom, cf eq.\ref{eq:vd}) for power-law models,
     using the red and blue (blue curve, marked R-B), red and
     intermediate (purple, R-I) or intermediate and blue (yellow, I-B)
     sub-populations.}}
 \label{fig:PLmodel}
 \end{figure}
 \begin{table}
 \begin{center}
 \renewcommand{\tabcolsep}{0.2cm}
 \renewcommand{\arraystretch}{0.5}
 \begin{tabular}{| l | c | c | c |}
 \multicolumn{4}{c}{} \\
         \hline 
 model & ${\hat \sigma}_{\rm b}$ & ${\hat \sigma}_{\rm i}$ & ${\hat \sigma}_{\rm r}$ \\  
   & (km/s) & (km/s) & (km/s) \\  
         \hline 
 broken power law (eq~\ref{eq:break}) & $332\pm15$ & $318\pm16$ & $296\pm25$ \\
 luminous \& dark (eq~\ref{eq:lumdark}) & $345\pm20$ & $290\pm22$ &  $316\pm35$\\
     \hline
 \end{tabular}
 \caption[GCs in M87: global velocity dispersions from different
   dynamical models]{\small{Inferred velocity dispersions, obtained by
     marginalising the posterior $\mathcal{L}$ over the photometric
     parameters and normalisations $\Sigma_{\rm r/b},\Sigma_{\rm i/b}$
     of central surface densities.}}
 \label{tab:vdispdyndec}
 \end{center}
 \end{table}

 \subsection{Scale-free Total Density}

 Given a model $\rho_{\rm tot}(r)$ for the total density, all
 population must satisfy the projected virial theorem
 simultaneously
 \begin{equation}
 \langle\sigma^{2}_{\rm p}\rangle_{j}=\frac{16\pi G}{3 L_{{\rm
       tot},j}}\int_{0}^{\infty}y\Sigma_{j}(y)\int_{0}^{y}\frac{r^{2}\rho_{\rm
     tot}(r)\mathrm{d}r}{\sqrt{y^{2}-r^{2}}}\ \mathrm{d}y 
 \label{eq:pvt}
 \end{equation}
 ~\citep[see e.g.,][and Paper I]{ae12a,am13}.
 Here, $\langle\sigma^{2}_{\rm p}\rangle_{j}$ is the average velocity
 second moment and $\Sigma_{j}(R)$ is the surface density profile of
 the $j-$th population, whilst $L_{{\rm tot},j}$ is the total
 luminosity of the population
 $2\pi\int_{0}^{\infty}R\Sigma_{j}(R)\mathrm{d}R.$

 As a first pass, let us begin with a scale-free total
 density. Although an over-simplification, the scale-free
 approximation has often led to useful
 insights in the past~\citep[c.f.,][]{chu10,ae12a}. The total density is given by
 \begin{equation}
 \rho_{\rm tot}(r)=\rho_{0}(r/r_{0})^{-\gamma}\ .
 \label{eq:PLdens}
 \end{equation}
 This has two independent quantities, the exponent $\gamma$ and a
 normalisation $\rho_{0}.$ Given a reference radius $R_{\rm d},$ it is convenient
 to define the velocity parameter
 \begin{equation}
 \vd =\ v_{\rm c}(R_{\rm d}) =\sqrt{\frac{4\pi
     G\rho_{0}R_{\rm d}^{2-\gamma}r_{0}^{\gamma}}{3-\gamma}}\ ,
 \label{eq:vd}
 \end{equation}
 which is the circular velocity at the radius $R_{\rm d}.$ When
 $\gamma=2$ (flat rotation curve), the dependence on $R_{\rm d}$
 vanishes and the average velocity dispersion for all sub-populations
 is simply $\vd/\sqrt{3}$, as can be deduced from eq.~(\ref{eq:pvt}).
 The scale $R_{\rm d}$ can be chosen arbitrarily. It just corresponds
 to the reference radius for mass measurements. Given our data, we have
 used $R_{\rm d}=10^{3}$ arcseconds for convenience.

 From the virial theorem applied to the $j-$th population, we obtain a
 predicted velocity dispersion ${\hat \sigma}_{j}(v_{\rm d},\gamma)$,
 averaged over the population's profile as in eq.~(\ref{eq:pvt}). We show
 its dependence on the two model parameters explicitly, but it also
 depends on the two photometric parameters $(R_{{\rm e},j},n_j).$
 These, in turn, come from the partition into three populations and
 every choice of the photometric parameters is weighted with the
 likelihood $\mathcal{L}$ of the decomposition. Then, we can quantify
 how the pair $(v_{\rm d},\gamma)$ provides a good fit to the
 kinematics of the $j-$th population as
 \begin{equation}
 L_{j}(\vd, \gamma)\propto \mathcal{L}(R_{{\rm
     e},j},n_{j}...)\times\mathrm{e}^{-({\hat
     \sigma}_{j}(\vd,\gamma)-\sigma_{j})^{2}/(2\delta\sigma_{j}^{2})}\
 \end{equation}
 marginalized over the other parameters ($R_{\rm e},n$ etc).  

 As a consequence, each \textit{pair} of sub-populations yields the
 scale-free density profile in the range spanned between the two
 effective radii considered. Let us recall that the enclosed mass in
 this class of models is $M(r)\propto r^{3-\gamma}.$ Then, from
 Fig.~\ref{fig:mest}, we can expect a steeper power-law density profile
 (i.e. higher $\gamma$) between the reddest and intermediate GCs and a
 shallower one between the intermediate and bluest population.
 Fig.~\ref{fig:PLmodel} shows the posterior distribution of the density
 exponent and velocity parameter $\vd$ when different pairs of
 populations are considered.  In order to make the comparison with
 measured velocity dispersions easier, we have displayed the
 distribution of $\vd/\sqrt{3}.$ The density profile is steeper between
 the red and intermediate-colour GCs and it becomes shallower in the
 region probed by the blue component, as expected.

 The scale-free results exhibit an interesting behaviour, which helps
 deepen our insight. The averaged velocity dispersion decreases or
 increases with $R_{\rm e}$ depending on whether $\gamma$ is larger or
 smaller than $2.$ {\it In other words, a dynamically hotter
   sub-population has a larger effective radius if the potential is
   shallower than isothermal, and a smaller effective radius if the
   potential is steeper than isothermal.}  This is nothing more than
 the dimensional scaling $\sigma\propto v_{c}(R_{e})$ with the proper
 coefficients given by eq.~(\ref{eq:pvt}).  The power-law exponents in
 Fig.~\ref{fig:PLmodel} agree with this picture. More precisely,
 $\sigma_{\rm r}>\sigma_{\rm i}$ corresponds to $\gamma>2$ in the
 region spanned by the red and intermediate-colour populations;
 conversely, $\gamma<2$ between the intermediate and blue GCs matches
 $\sigma_{\rm b}>\sigma_{\rm i}$.

 \begin{figure*}
  \includegraphics[width=\textwidth]{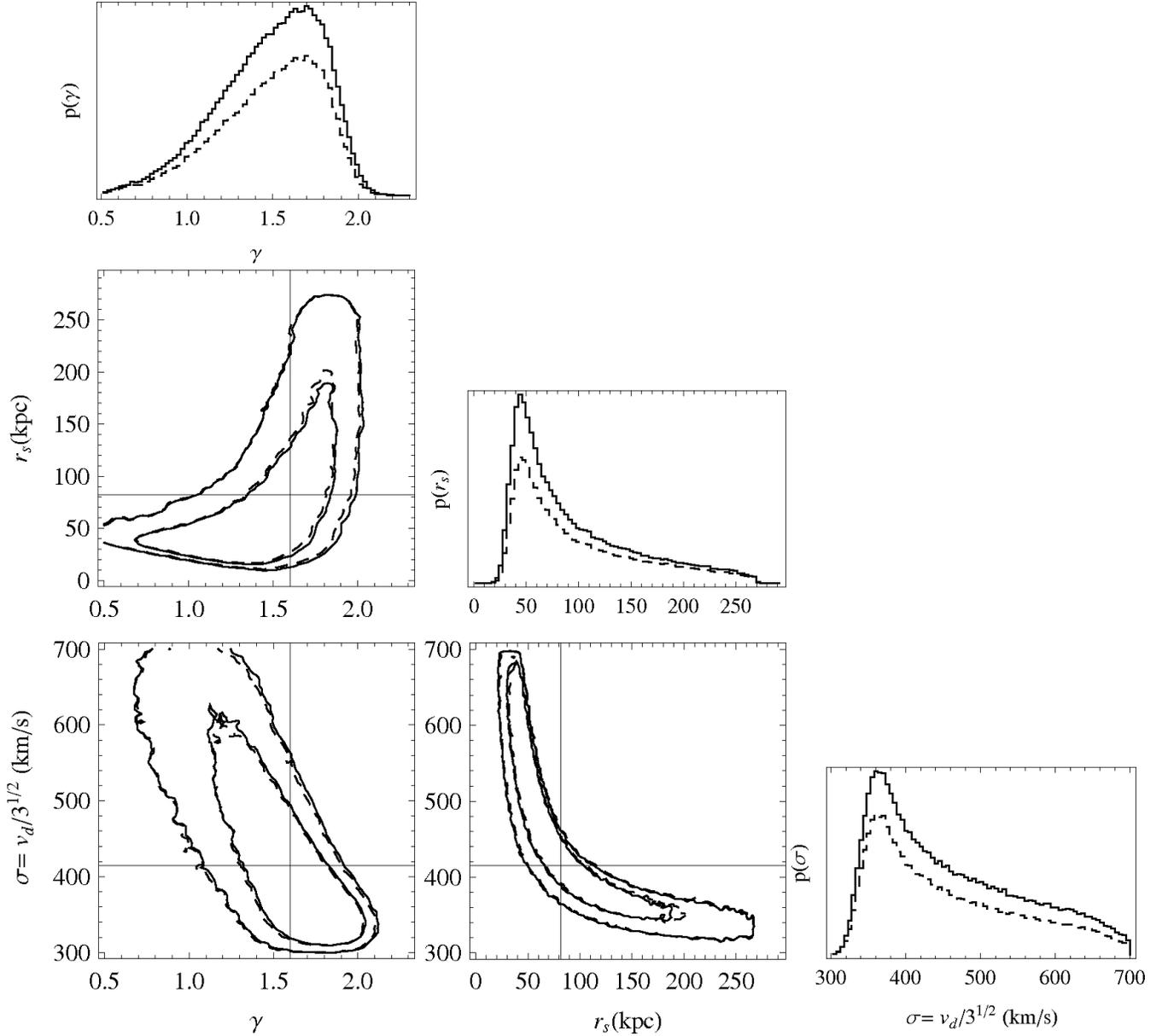}
  \caption[GCs in M87: broken power-law total density]{\small{
 Posterior on the parameters of the broken-power-law
      model of eq~(\ref{eq:break}). We show the $68\%$
      and $95\%$ confidence regions of the marginalised
      posterior distributions in the space of power-law slope $\gamma$,
      logarithm of the scale radius $\log_{10}(r_{\rm s}\ (\mathrm{arcsec}))$
      and velocity dispersion $\vd/\sqrt{3}$ (in km/s), besides
      the resulting distributions of these three parameters.
      Full lines: contours from the full parameter space, including
      models with $R_{\rm e,i}>R_{\rm e,b}$. Dashed lines: contours from
      the marginalisation over parameter space with $R_{\rm
      e,i}<R_{\rm e,b}$. The inference on dynamical parameters is then
      independent of the morphological ($R_{e},n$) ones.
      The axes in the correlation plots intersect
      at the best-fitting values (cf.Sect.4.3), which do not always
      correspond to the marginalised ones.}}
 \label{fig:bPLpost}
 \end{figure*}

 \subsection{Broken Power-law Density}

 The shallower behaviour of $\rho(r)$ at large radii could be a
 spurious effect, since a single power-law may not be an adequate
 description over the whole extent of the tracers. A similar
 phenomenon was studied in \citet{aae13}, where broken power-law
 densities
 \begin{equation}
 \rho(r)=\rho_{0}\frac{(r/r_{0})^{-\gamma}}{\left(1+r^{2}/r_{\rm s}^{2}\right)^{(3-\gamma)/2}}
 \label{eq:break}
 \end{equation}
 were analysed by means of scale-free models. A broken power-law
 density introduces an explicit length-scale, so that there are now
 three independent parameters. If dynamical measurements extend out to
 the break radius $r_{\rm s}$, a scale-free model $\rho\propto
 r^{-\gamma_{\mathrm{g}}}$ will strongly bias the estimated exponent
 towards $\gamma_{\mathrm{g}}=2$ and produce a non-monotonic
 dependence of $\gamma_{\mathrm{g}}$ on the true exponent $\gamma.$

 Since we have three sub-populations at our disposal, we can demand
 that they satisfy the virial theorem simultaneously and infer the
 likelihood distribution for all the three parameters, $\rho_0,\gamma$
 and $r_{\rm s}.$ Instead of $\rho_0$, we will continue to work with
 $\vd$ as defined in eq.~(\ref{eq:vd}).  This is no longer the
 circular velocity at $R_{\rm d}$ (unless $r_{\rm s}\gg R_{\rm d}$),
 but it makes the comparison with scale-free models easier. Once
 again, eq.~(\ref{eq:pvt}) yields three velocity dispersions ${\hat
   \sigma}_{j}$ as functions of the two photometric parameters
 $(R_{{\rm e},j},n_{j})$ and of the three dynamical parameters
 $(\vd,\gamma,r_{\rm s}).$

 In this case, we can incorporate the dynamical model within the
 sub-population partition, by using the three ${\hat \sigma}_{j}$
 instead of $\sigma_{j}$ in eq.~(\ref{eq:like3pop}) for the likelihood
 $\mathcal{L}.$ This, in turn, is now a function of the three
 dynamical parameters $(\vd, \gamma,r_{\rm s})$, in addition to the
 colour $(\langle c\rangle=\langle g-i\rangle, \Delta c)$ and photometric parameters
 $(R_{{\rm e}}, n)$ for each of the three populations (red,
 intermediate and blue) plus the two normalisations, $\Sigma_{\rm
   r/b}=\Sigma_{\rm r}(0)/\Sigma_{\rm b}(0)$ and $\Sigma_{\rm
   i/b}=\Sigma_{\rm i}(0)/\Sigma_{\rm b}(0).$

 The marginalised likelihood in the colour parameters and fractions
 $(\Sigma_{\rm r/b},\Sigma_{\rm i/b})$ is the same as for the
 model-independent partition of Section 4.2. However, two differences
 arise in the correlations between effective radii and in the
 posteriors inferred on the velocity dispersions $({\hat \sigma}_{\rm
   b},{\hat \sigma}_{\rm r},{\hat \sigma}_{\rm i})$. First, the
 model-independent partition privileges a strict ordering $R_{\rm
   e,b}>R_{\rm e,i}$ on the effective radii, whereas this dynamical
 decomposition allows for the possibility of reverse ordering.  In
 particular, the posterior probability $P(R_{\rm e,i}>R_{\rm e,b})$ is
 approximately $40 \%.$ Secondly, there is a mismatch
 between the optimal dynamical parameters and the peak of the
 marginalised posterior distribution. The likelihood $\mathcal{L}$ is
 maximised for $\gamma=1.52,$ $\vd=455\sqrt{3}$ and $r_{\rm s}=0.26$
 degrees, and corresponding velocity dispersions ${\hat \sigma}_{\rm
   b}=344,$ ${\hat \sigma}_{\rm r}=313,$ ${\hat \sigma}_{\rm i}=287$
 kms$^{-1}$, which are compatible with the findings of Section 4.2.  On
 the other hand, the \textit{marginalised} posterior distribution in
 ${\hat \sigma}_{j}$ peaks at ${\hat \sigma}_{\rm b}=332,$ ${\hat
   \sigma}_{\rm r}=296,$ ${\hat \sigma}_{\rm i}=318$ kms$^{-1}$. If the
 condition $R_{\rm e,b}>R_{\rm e,i}$ is enforced, there is no way of
 obtaining the right ordering in the velocity dispersions.  In other
 words, the broken power-law model requires the intermediate-colour
 population to be more extended than the blue one, in order to
 reproduce the best-fitting velocity dispersions. Alternatively, if the
 blue component is broader than the intermediate one, the velocity
 dispersions are reduced (for the blue and red populations) or
 increased (for the intermediate population) with respect to the
 outcome of the model-independent partition.

 Fig.~\ref{fig:bPLpost} shows the marginalised posterior distribution
 in the three dynamical parameters $(\gamma,\sigma,r_{s}).$
 The triple $(\vd, \gamma,r_{\rm s})$ is correlated with the photometric
 parameters, a fact that is reflected in the peculiar behaviour of the inferred
 velocity dispersions explained above. 

 \begin{figure*}
  \includegraphics[width=\textwidth]{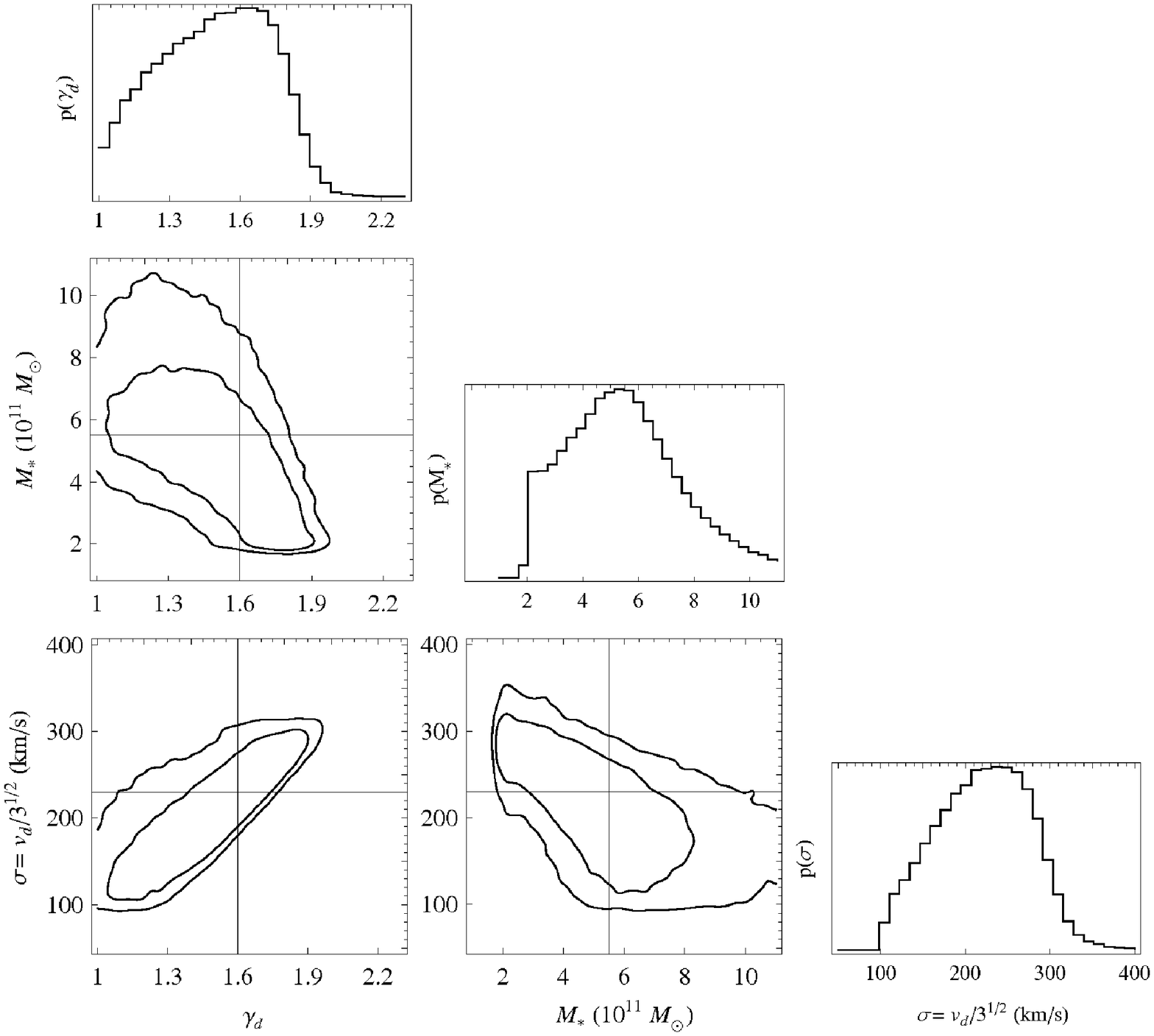}
 \caption[GCs in M87: luminous/dark decomposition]{\small{ Posterior
     on the parameters in the luminous and dark decomposition given in
     eq~(\ref{eq:lumdark}). We show the $68\%$ and $95\%$ confidence regions
     of the marginalised posterior in the space of of stellar mass $M_{\star}\ $(in units of
     $10^{11}M_{\odot})$, dark halo power-law slope $\gammad$, and
     velocity dispersion $\vd/\sqrt{3}$ (in kms$^{-1}$).  The axes
     intersect at the best-fitting values, which roughly correspond to
     the peaks of the marginalised distributions.  }}
 \label{fig:LDdec}
 \end{figure*}

 \subsection{Luminous and Dark Matter Decomposition}

 \label{sect:LDdec}

 A more faithful description of the system may consist of a
 luminous and dark matter decomposition. The luminous component may
 contribute to the steeper inner density, whilst the DM density may
 indeed be shallower than $r^{-2}$. In particular, we can consider the
 following model:
 \begin{equation}
 \rho_{tot}(r)=\frac{M_{\star}}{2\pi^{2}
   r_{\star}r^{2}(1 +
   r^{2}/r_{\star}^{2})}+\rho_{0,\rm d}(r/r_{0,\rm d})^{-\gammad}\ .
 \label{eq:lumdark}
 \end{equation}
 The luminous profile (proportional to $M_{\star}$) is similar to the
 commonly used Jaffe (1983) profile, but is more convenient
 computationally. The radius $r_{\star}$ is chosen so that the
 effective radius is equal to that of the starlight,
 $R_{e,\star}=0.02$ degrees \citep[see][and references
   therein]{har09}.  The free parameters are the stellar mass
 $M_{\star},$ the DM exponent $\gammad$ and the DM circular
 velocity at $R_{\rm d}$ (which we choose without loss of generality
 as $1000$ arcsec):
 \begin{equation}
 v_{\rm d}=\sqrt{\frac{4\pi G\rho_{\rm 0,d} R_{\rm d}^{2-\gammad} r_{0,\rm d}^{\gammad}}{3-\gammad}}\ ,
 \end{equation}
 plus the photometric and colour parameters.

 The virial theorem gives three velocity dispersions ${\hat
   \sigma}_{\rm b,r,i}$ as a function of $(M_{\star}, v_{\rm d},
 \gammad)$ and the photometric parameters, which are used in the
 maximum likelihood method via eq.~(\ref{eq:like3pop}). Now the
 posterior distribution reproduces the same features of the
 model-independent partition, both in terms of effective radii
 ($R_{\rm e,b}>R_{\rm e,i}$) and velocity dispersions, which are
 listed in Table \ref{tab:vdispdyndec}. Figure \ref{fig:LDdec} shows
 the marginalised likelihood in terms of the dynamical
 parameters. Within this model, the posterior has a peak at
 $M_{\star}= 5.5\times 10^{11}M_{\odot},$ $\gammad = 1.6$ and $ \vd=
 230\sqrt{3}.$ The $1\sigma$ confidence intervals for the parameters
 are: $1.31<\gammad<1.76,$ $3.0<M_{\star}(10^{11}M_{\odot})<6.9,$ and
 $170<\vd/\sqrt{3}\,\rm{km/s} <280.$ Then, the underlying profile is
 reliably described as the superposition of simple luminous and dark
 components, although the inferred dynamical parameters inherit the
 large uncertainty from the photometric parameters. At very low values
 of $M_{\star},$ the solution with $\gammad=2$ and $\vd=350\sqrt{3}
 \approx 600$ kms$^{-1}$ is available. This conspiracy between
 luminous and dark matter to produce an overall $\rho\propto r^{-2}$
 profile has been observed in other cases \citep{tk04,hum10,rem13},

Finally, it is interesting to compare the best-fitting dark matter
 masses to our earlier estimates in Table~\ref{tab:masses}.
 Integrating our dark matter density law out to $0.46^\circ \approx 135$
 kpc, the dark matter mass is $8.0 \times 10^{12}M_\odot$. Added to the
 luminous mass of $5.5\times 10^{11}M_{\odot},$ gives a total mass of $8.6 \times
10^{12}M_\odot$, reassuringly close to the value of $9 \times
10^{12} M_\odot$ given by the cruder estimates of Sect. 4.1}. The total mass
 at $135$ kpc, inferred from the likelihood on the dynamical parameters, has
 sizeable uncertainties, namely
 $M_{tot}(135 \rm kpc)=8.0^{+1.0}_{-4.0}\times 10^{13} M_{\odot}.$

\subsection{Evidence for DM Contraction}\label{sect:DMcontr}

The model with luminous and dark components as in eq.~(\ref{eq:lumdark})
 predicts kinematic and photometric properties compatible with the
model-independent results of Section~\ref{sect:3pops}.  At the same
time, it yields information on the best fitting power-law slope of the
dark matter. Although the posterior distribution of $\gammad$ in
Fig.~\ref{fig:LDdec} is broad, the peak of the distribution lies at
$\gammad \approx 1.6$. This suggests that the underlying DM density
may be steeper than $r^{-1}$, which is predicted at small radii by
cosmological DM-only simulations \citep[][hereafter NFW]{dub91,nfw96}.
However, this result may be the outcome of a model, where the DM halo
has a scale-free density.  Perhaps the DM density is more
appropriately described by a broken power-law similar to
eq.~(\ref{eq:break}), which is a simple generalisation of the NFW
profile.

We can test whether this is a mere consequence of the adopted model by
allowing the DM density to be a broken power-law, parameterized by an
inner exponent $\gammad,$ a normalisation $\vd$ and a break-radius
$r_{\rm s}$ as in eq.(22). Within the virial method, we can use these as free
parameters in the decomposition into three populations, provided the
luminous mass $M_{\star}$ is kept fixed. The procedure can be repeated
for different fixed values of $M_{\star},$ to quantify the change in
the inferred $\gammad,\vd$ and $r_{\rm s}.$ The posterior distribution
(marginalised over colour and photometric parameters) has the same
width regardless of the chosen value for $M_{\star},$ whereas the
maximum likelihood values vary as

\begin{eqnarray}
\gammad(M_{\star}) &\approx&
1.575-0.075\left(\frac{M_{\star}}{10^{11}M_{\odot}}-5.4\right)\ ,\nonumber\\ \vd
/
\sqrt{3}\ &\approx&297.5-12.5\left(\frac{M_{\star}}{10^{11}M_{\odot}}-5.4\right)\ {\rm
  km/s},\\ r_{\rm s}\ &\approx&
590-67\left(\frac{M_{\star}}{10^{11}M_{\odot}}-5.4\right)\ {\rm
  arcsec}.\nonumber
\end{eqnarray}
The uncertainties are symmetric for $\gammad$ and $\vd/\sqrt{3}$ and
amount to $0.2$ and $45$ kms$^{-1}$ respectively, whereas $r_{\rm s}$
has a very skew posterior distribution around the peak value. We can
marginalise the inner exponent over $M_{\star},$ by considering a sum
of Gaussians with mean $\gamma_{d}(M_{\star})$ and dispersion $\delta
\gamma=0.2$ and weighing $M_{\star}$ with the virial likelihood
(Section~\ref{sect:LDdec} and Fig.~\ref{fig:LDdec}). The resulting
exponent is $\overline{\gammad}=1.55\pm0.25,$ which is still
appreciably different from the simple NFW prediction $\gammad = 1.$
Furthermore, a decomposition relying strictly on a NFW halo gives
inversions in the effective radii and velocities dispersions similarly
to that found in Section 4.3. These findigs lend decisive support in favour
 of DM contraction, with a DM density exponent $\gamma_{d}\approx 1.6.$

\begin{figure*}
 \includegraphics[width=\textwidth]{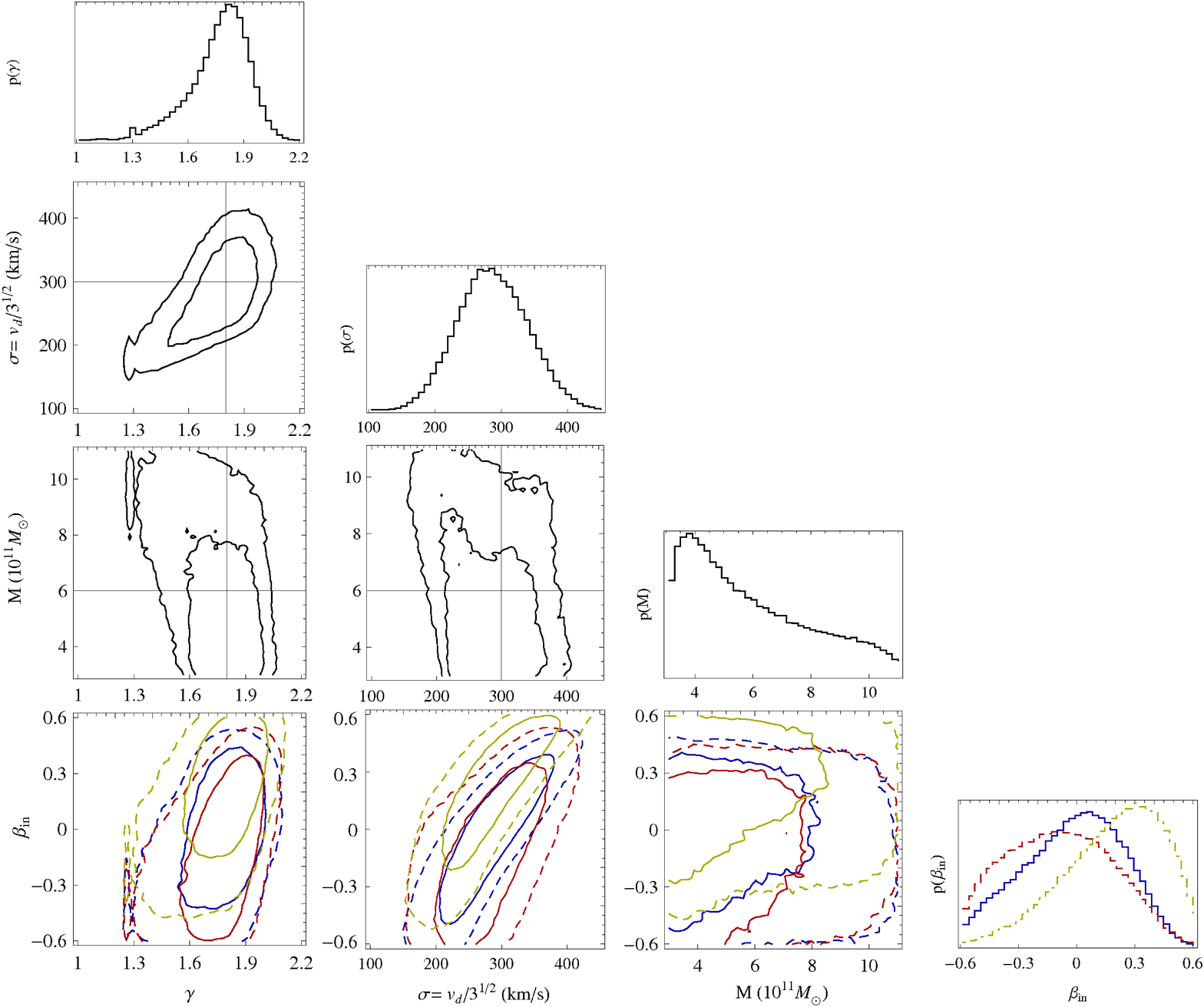}
\caption[GCs in M87: Jeans analysis]{\small{ Posterior on the
    parameters in the luminous and dark decomposition, together with
    the inner anisotropy, as inferred from the Jeans equations
    (Section~\ref{sect:Jeans} and Paper I). We show the $68\%$
    and $95\%$ confidence regions in parameter space, together with
    the marginalised posterior distribution on the single parameters.
    Different colours indicate the parameters of different populations:
    blue full lines (red dashed lines, and yellow dot-dashed lines)
    stand for the blue (red, and intermediate) populations. The axes
    intersect at the best-fitting values; the best-fitting $M_{\star}$ does not
      correspond to the peak of the marginalised likelihood, because
      of the mass-anisotropy degeneracy, whereas $\gamma$ and $\sigma$
      follow the same covariance as in the virial analysis.  }}
\label{fig:Jeans}
\end{figure*}

\section{Jeans analysis}\label{sect:Jeans}

In the previous sections, the dynamics of the system has been explored
through the projected virial theorem, which is derived through integration of
the Jeans equations over configuration space.

If instead the Jeans equations are employed directly, then the
sub-populations are described by the previous parameters
 $(R_{\rm{e},j}, n_{j},...)$ and also by their anisotropy profiles
\begin{equation}
\beta(r)=1-\frac{\langle v_{\rm t}^{2}\rangle}{2\langle v_{r}^{2}\rangle}.
\end{equation}
Here, $\langle v_{r}^{2}\rangle$ and $\langle v_{\rm t}^{2} \rangle$
are the radial and tangential velocity second moment, while $r$ is the
spherical polar radius.

 In our analysis, we will use
\begin{equation}
\beta(r)={\bin r_{\rm a}^2 + \bout r^2 \over r_{\rm a}^2 + r^2}
\end{equation}
so that the inner anisotropy is $\bin$ and the outer anisotropy
$\bout$, whilst $r_{\rm a}$ is a transition radius. 
 Here, we use eq (16) of Paper I to calculate the velocity
dispersions averaged within radial annuli, and thus build kinematic
profiles, which are then used to separate the populations similarly
 to the procedure followed in the previous Sections.

As already noted in Section~\ref{sect:3pops}, the use of parameterised
kinematic profiles, such as in eq~(\ref{eq:sprofblue}), does not
improve the inference on average velocity dispersions, which are
linked to the mass model via the virial theorem. Thus, we can expect
that the Jeans analysis will provide information on the velocity
anisotropy profile, but no significant additional constraints on the
mass model.

\subsection{Dynamical parameters}

The results of the previous sections suggest that we restrict
attention to the two-component mass-model of eq.~(\ref{eq:lumdark}),
which is both realistic (Section \ref{sect:LDdec}) and reliable
(Section \ref{sect:DMcontr}).

Once the likelihood is marginalised over the colour and photometric
parameters, the posterior in the dynamical parameters
$(M_{\star},\vd,\gamma_{\rm d})$ and the anisotropies is left.  In
Fig.~\ref{fig:Jeans}, we show the inference on the inner
anisotropies $\bin$ of the three populations and on the dynamical
parameters. The
mass-anisotropy degeneracy is evident in the correlation between
$\bin$ and the DM parameters $(\gamma,\vd),$ for each population.
 There is a weaker correlation between this set of
parameters and the remaining ones $(\bout,r_{\rm a}).$ The
posterior distribution of the dynamical parameters is similar to what
we found in Section \ref{sect:LDdec}.

Compared to the virial results, the uncertainties on the DM parameters
are slightly smaller but comparable. The likelihood profile is very
shallow in $M_{\star}$ and the marginalised posterior inherits its
profile from the behaviour of the likelihood in the other parameters.
For example, at small values of $M_{\star}$ a wider range of
anisotropies is allowed, giving a larger value for the integrated
likelihood towards the low-mass end. This is another effect of the
mass-anisotropy degeneracy. Having chosen a particular model for
$\beta(r),$ Jeans methods will privilege a narrower range of dynamical
parameters, whereas the virial results account for the whole
generality of anisotropy models that could be chosen.

\begin{figure}
\centering
 \includegraphics[width=0.35\textwidth]{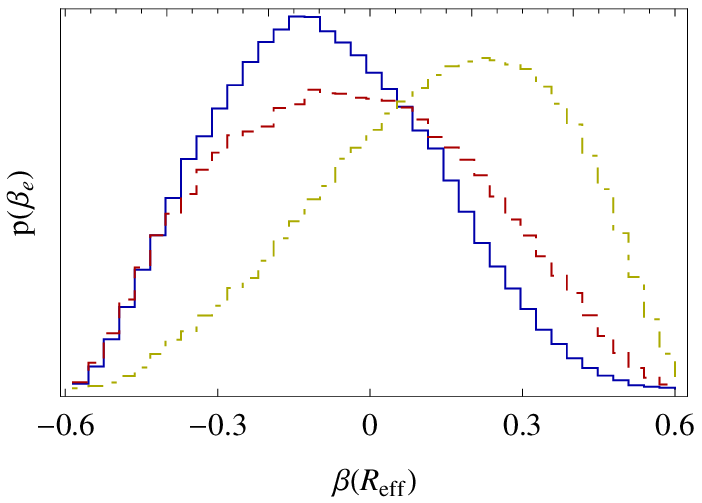}\\
 \includegraphics[width=0.35\textwidth]{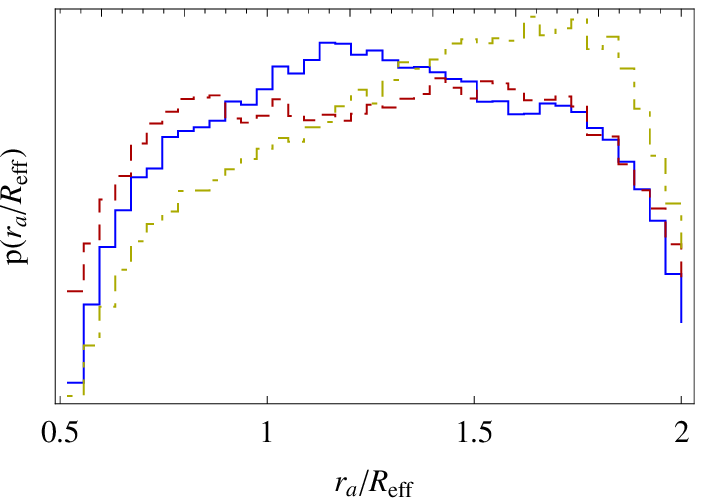}\\
\caption[GCs in M87: anisotropies]{\small{
 Marginalised posterior distribution on the anisotropy parameter at the effective radius
    $\beta_{\rm e}=\beta(R_{\rm e})$ (top panel) and the ratio of anisotropy radius to
    the effective radius $r_{\rm a}/R_{\rm e}$ (bottom panel). Colour
    coding as before: blue solid lines (red dashed lines, and yellow
    dot-dashed lines) stand for the blue (red, and intermediate)
    populations.}}
\label{fig:Beta}
\end{figure}

\subsection{Anisotropies}\label{sect:ani}

The inferences on outer anisotropies and anisotropy radii are shown in
Fig.~\ref{fig:Beta}.  The value of $r_{\rm a}/R_{\rm e}$ is not well
constrained, the only reliable information being a lower limit $r_{\rm
  a}\gtrsim R_{\rm e}/2$ \citep[cf][for a similar case]{son12}. A
narrower distribution is obtained if the anisotropy at the effective
radius $\beta(R_{\rm e})$ is considered, which is shown in the last
panel.

A glance at the inferred anisotropy profiles (figures 8 and 9) shows some
interesting features of the system. The red GCs are on slightly
tangential orbits, with nearly constant anisotropy.  This may be simply
consistent with the supposition that GCs at sufficiently small
 distances from M87 are tidally shredded at pericentric passage
 unless their orbits have sufficient angular momentum \citep[cf][]{web13}.
 This creates a \textit{loss-cone} \citep{bah76} in velocity space,
 whose aperture decreases with the distance from the center
 (as $r^{-2}$ for a flat rotation curve). The intermediate-colour GCs,
 which orbit at larger radii, have a mildly radial
 anisotropy, $\beta(r)\approx0.3.$

On the other hand, the blue GCs have approximately isotropic orbits at
smaller radii and a mildly tangential velocity dispersion tensor at
larger distances. This behaviour cannot be explained with arguments of
tidal disruption, since the loss-cone in velocity space is already
small enough for the intermediate-colour GCs to survive on mildly
radial orbits. There is, however, another dynamical phenomenon that
can produce such an anisotropy profile, which is related to the
accretion process of the central object. If M87 has accreted mass on
sufficiently slow time-scales, then radial orbits are dragged towards
the center more efficiently than tangential ones
\citep{goo84,lee89,cip94}, which then contributes to a net tangential
anisotropy in the outer parts. This scenario of (approximately)
adiabatic contraction would also agree with the enhanced DM exponent,
$\gammad\approx1.6,$ significantly larger than unity (Sections 4.4, 4.5).

Orbital time-scales are shorter at smaller distances. Then, if the
intermediate-colour GCs are coeval with the blue ones, they must have
experienced a similar phenomenon. The inferred positive value of
$\beta_{\rm out}$ for this population suggests that they have been
falling onto M87 on preferentially radial orbits, differently from the
more distant and bluer GCs. 

\section{Conclusions}

The giant elliptical galaxy M87 is surrounded by one of the largest
known populations of globular clusters (GCs), exceeding $10^4$ in
total \citep{tam06}. This extensive swarm of GCs offers a wealth of information
regarding M87's mass profile, together with tantalizing evidence about
mechanisms of its formation and evolution. We have been able to answer
some questions, thanks to the availability of a new data set of
positions, colours and velocities with the largest radial extent and
best kinematic accuracy  to date (Strader et al 2011, or S11).

\subsection{Multiple populations}

The colour distribution and kinematics of the GCs at different radii
suggest a partition into multiple sub-populations (see, for example,
Harris 2009, S11 and references therein). Here, we have shown that a
separation into three components is statistically preferred over ones
into two or four populations for M87's GC system. To separate the
three components (blue, intermediate and red GCs), we have exploited a
maximum-likelihood method in which the information from position,
colour and velocity is used jointly.  The average velocity dispersion
and the colour-distribution parameters of each population are robustly
determined, whereas the S{\'e}rsic indices and half-mass radii are
affected by sizeable uncertainties (see for example,
Table~\ref{tab:3pop}).  The uncertainty in turn is a consequence of
the limited number ($N \approx 420$) of bona fide GCs with high
quality spectroscopic data.

A case for more than two GC populations has been made also in studies of other
 galaxies, most notably NGC 4365 \citep{bro05}.
 The same system has been reanalysed by \citet{blo12}, where the
 colours of the three sub-populations have been linked to their different
 kinematics and viable formation scenarios. Here, we have shown how the
 dynamics (density profile and anisotropies) can be used, within a robust
 Bayesian analysis, to optimally decompose the GC system in sub-populations
 as well as compare formation scenarios with observational properties.

\subsection{Masses}
\begin{figure}
\centering
 \includegraphics[width=0.47\textwidth]{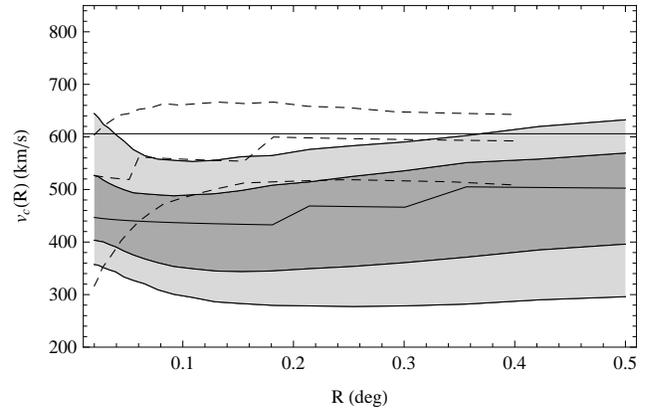}\\
\caption[GCs in M87: circular velocity]{\small{
     Circular velocity curve as inferred from M87's GCs.
     The grey-shaded regions mark $68\%$ and $95\%$ confidence
     levels for the model with luminous and dark matter (Sect.4.4).
     The dashed lines show the average and $68\%$ confidence level
     inferred from the dynamical model with a broken power law (Sect.4.3),
     which however is discarded after comparison with the model-independent
     partition of Sect.3.2 and is shown here just for the sake of completeness.
     The horizontal axis at $v_{c}=606$ km/s corresponds to the case
     of a flat circular velocity curve, $\rho_{tot}\propto r^{-2}$.}}
\label{fig:vcirc}
\end{figure}

\noindent
The kinematics of the GCs are a direct probe of the total mass distribution
of M87. Each sub-population with its own global velocity
dispersion and photometric parameters has its own virial pinch
radius (eq.\ref{eq:paperi},\ref{eq:paperia} and Paper I).
 Therefore, the red, intermediate and blue GC populations
provide us with three enclosed masses at three different pinch radii
(see Table~\ref{tab:masses}). The mass estimate at the pinch radius
of the red GCs is corroborated by the results of more elaborate
techniques applied to the stellar and GC populations~\citep{mur11}.  The
method is robust, as it is derived as a blind average over different
possible halo mass models.  The largest uncertainty is in the
determination of the effective radius, and so the covariance between
pinch radius and enclosed mass follows the scaling
$M(R_{\star})\propto R_{\star}$ inherited from
eq~(\ref{eq:paperi}). This does not imply that the underlying mass
distribution corresponds to that of a flat rotation curve.  Indeed,
there is a danger that the uncertainties scatter solutions along the
locus of flat rotation curve models, which really amounts to fitting
the covariance of the data rather than the underlying mass profile.

The mass profile can also be constrained by using dynamical modelling
to obtain the velocity dispersions, which are then used to separate the
three components. The first step consists of applying the Projected
Virial Theorem (eq.~\ref{eq:pvt}).  This has the advantage of
involving the surface-density averaged (i.e. global) velocity dispersions only,
thus avoiding systematic uncertainties arising from the
mass-anisotropy degeneracy. By comparing the results of the separation
relying on the virial theorem with those of the model-independent
decomposition, we have been able to validate the luminous and dark
matter model of eq~(\ref{eq:lumdark}), placing constraints on the
luminous mass, dark matter (DM) density exponent and normalisation.
The luminous and dark masses at different locations, together with their
 uncertainties, are discussed in Sections 4.1, 4.3 and 4.4.
 Figure \ref{fig:vcirc} displays these results in terms of the inferred
 circular velocity at different locations.
 
The inferred density exponent is $\gamma\approx 1.6,$ if a scale-free
DM profile is used.  When a broken power-law model for the DM halo is
adopted, the inner exponent is still in the same range ($\gamma\approx
1.55$). This is consistent with the findings of \citet{son12} for an
early-type galaxy of similar mass and can be interpreted as evidence
for DM contraction.  The scale-radius $r_{\rm s}\approx600$ arcsec where the DM density
profile changes from $\sim r^{-\gamma}$ to $\sim r^{-3}$ is somewhat
smaller than the effective radius of the intermediate-colour
population, but appreciably larger (or smaller) than the effective
radius of the red (or blue) component.

\subsection{Comparison with X-ray Studies}

Studies in the literature compare the masses measured from the hot
X-ray gas to those inferred from GC kinematics \citep{mur11}. The
X-ray masses are typically lower than the ones from GCs at small radii
and larger at large radii. The discrepancy at small distances has been
alleviated by invoking some amount of non-thermal support for the gas,
but the disagreement at large distances would still hold
\citep{hum13}. It is fair to say that previous studies relied on
velocity dispersions that were appreciably overestimated, so the role
of non-thermal motions is far from clear. The complex and asymmetric
X-ray photometry of the gas calls for further caution in deriving
masses from the X-ray gas based on assumptions such as spherical
symmetry.

Another worry is the assumption that the gas is exactly in hydrostatic
balance.  If $u(r)$ is the radial velocity of the gas, then from the
Euler equation for a fluid flow with pressure $p$ and density $\rho$

\begin{equation}
\partial_{t}u+u\frac{\partial u}{\partial r}\equiv\frac{\mathrm{d}u}{\mathrm{d}t}=-\frac{1}{\rho}\partial_{r}p-\frac{GM(r)}{r^{2}}\ ,
\end{equation}
we see that depending on how the accreted gas settles towards
hydrostatic equilibrium (i.e., positive or negative acceleration), the
underlying mass profile may be under or over-estimated. This aspect
has been discussed in detail for other systems \citep{cio04,pel06},
and there is evidence against hydrostatic equilibrium.  Given the high
virial mass of M87, it is plausible to assume that gas has been
accreting onto it in the hot mode~\citep{bir03}, thus slowly setting
towards hydrostatic balance after shocking at the virial radius. If
the gas has bounced off the central regions, as a consequence of
pressure build-up from the converging inflow, then it might be still
experiencing a negative acceleration (from positive to zero velocity)
and the mass from X-ray studies may be an underestimate. If at large
distances the gas is still passing from negative to zero velocity,
then the convective acceleration is positive and the X-ray mass may be
an overestimate.

\citet{das10} found that the X-ray velocity curve of M87 rises from 500
km/s at 6 kpc (the effective radius of the starlight) to 600
km/s at large radii. They suggest that these numbers are
susceptible to a 10 \% uncertainty due to systematic effects. This
takes them to values close to ours inferred from dynamical modelling,
which range from 500 km/s at 6 kpc to 530 km/s at large
radii.  We conclude that there is tendency for the X-ray masses to be slightly
overestimated at large radii, but that the discrepancy is within the
uncertainties.

\subsection{Orbits}

The results from the Jeans analysis suggest an interesting
distribution in velocity space. On the one hand, the GCs at small
radii are consistent with a loss-cone distribution function that
privileges tangential orbits, as GCs on radial orbits suffer tidal
disruption near pericentric passage. On the other hand, at large distances
the anisotropy profile of the blue GCs tends again towards mild
tangentiality, which cannot be explained by loss-cone arguments. A
more plausible scenario would invoke accretion of external material
onto M87 once the GCs are already in place.  The build-up of mass has
a different effect on GC orbits depending on their angular momenta. If
accretion proceeds slowly enough, this gives rise to a tangentially
biased velocity-dispersion tensor in the outer parts, together with a
contraction of the DM density profile at smaller distances.

Inference on the structure in velocity space has been possible within
the framework of the simple analysis presented here. This is based on
the Jeans equations and the approximation of spherical symmetry, which
is appropriate as a first step in the case of M87.  The use of more
refined modelling techniques, such as non-spherical Jeans modelling,
 orbit-based or made-to-measure, would be useful to encode additional
effects like possible flattening of the dark halo and more elaborate
velocity distributions. However, the analysis of \citet{AmE12} and the
sizeable statistical uncertainties on the surface-density parameters
suggest that more advanced techniques will have to wait until larger
data sets are collected.

\section*{Acknowledgments}
We are indebted to Jay Strader for his observational effort on the
 M87 GC system, which has resulted in a sound basis for this work.
 We thank Luca Ciotti, Daniel Wang, Cathie Clarke and Andy Fabian for
 helpful discussions on X-ray masses, and Vasily Belokurov for
 significant feedback on the manuscript. We thank the anonymous referee
 for a very detailed report, which helped improve the manuscript considerably.
 AA acknowledges financial support from the Science and Technology
 Facilities Council (STFC) and the Isaac Newton Trust. AJR and JPB were supported
 by National Science Foundation grants AST-0909237 and AST-1211995.

\label{lastpage}

\end{document}